\documentclass[a4paper,10pt,onecolumn,openright,twoside]{article}

\usepackage[utf8]{inputenc} 
\usepackage[T1]{fontenc}      
\usepackage[english]{babel}  
\usepackage{caption}
\usepackage[round]{natbib}
\usepackage[top=2cm, bottom=2cm, left=2cm, right=2cm]{geometry} 
\usepackage{supertabular}
\usepackage{setspace}
\usepackage{color} 
\usepackage{lmodern} 
\usepackage{url} 
\usepackage{graphicx} 
\usepackage{amsmath}
\usepackage{amssymb}
\usepackage{mathrsfs}
\usepackage{eurosym}
\usepackage{lineno}
\usepackage{multirow}
\usepackage{multicol}
\usepackage[squaren, Gray, cdot]{SIunits}

\makeatletter
\newcommand\sixteen{\@setfontsize\sixteen{17pt}{6}}
\renewcommand{\maketitle}{\bgroup\setlength{\parindent}{2pt}
\begin{flushleft}
\sixteen\bfseries \@title
\medskip
\end{flushleft}
\textit{\@author}
\egroup}
\makeatother

\title{A fractal model for the electrical conductivity of water-saturated porous media during mineral precipitation-dissolution processes}

\author{Flore Rembert$\,^{1,\star}$, Damien Jougnot$\,^{1}$, Luis Guarracino$\,^{2}$ \\ \\
$^{1}\ $Sorbonne Universit\'e, CNRS, UMR 7619 METIS, FR-75005 Paris, France \\
$^{2}\ $CONICET, Faculdad de Ciencias Astron\'omicas y Geof\'isicas, Universidad Nacional de La Plata, Paseo del Bosque s/n, 1900 La Plata, Argentina \\ \\
$^{\star}\ $Corresponding author: flore.rembert@sorbonne-universite.fr}

\begin{document}

\maketitle

\begin{onehalfspace}

\paragraph{Highlights}
    \begin{itemize}
        \item[$\bullet$] A new electrical conductivity model is obtained from a fractal upscaling procedure
        \item[$\bullet$] The formation factor is obtained from microscale properties of the porous medium
        \item[$\bullet$] Transport properties are predicted from the electrical conductivity
        \item[$\bullet$] The model can reproduce dissolution and precipitation processes in carbonates
    \end{itemize}
    
\paragraph{Abstract}Precipitation and dissolution are prime processes in carbonate rocks and being able to monitor them is of major importance for aquifer and reservoir exploitation or environmental studies. Electrical conductivity is a physical property sensitive both to transport phenomena of porous media and to dissolution and precipitation processes. However, its quantitative use depends on the effectiveness of the petrophysical relationship to relate the electrical conductivity to hydrological properties of interest. In this work, we develop a new physically-based model to estimate the electrical conductivity by upscaling a microstructural description of water-saturated fractal porous media. This model is successfully compared to published data from both unconsolidated and consolidated samples, or during precipitation and dissolution numerical experiments. For the latter, we show that the permeability can be linked to the predicted electrical conductivity.


\paragraph{}
\begin{figure}[ht]
    \centering
    \includegraphics[width=300pt]{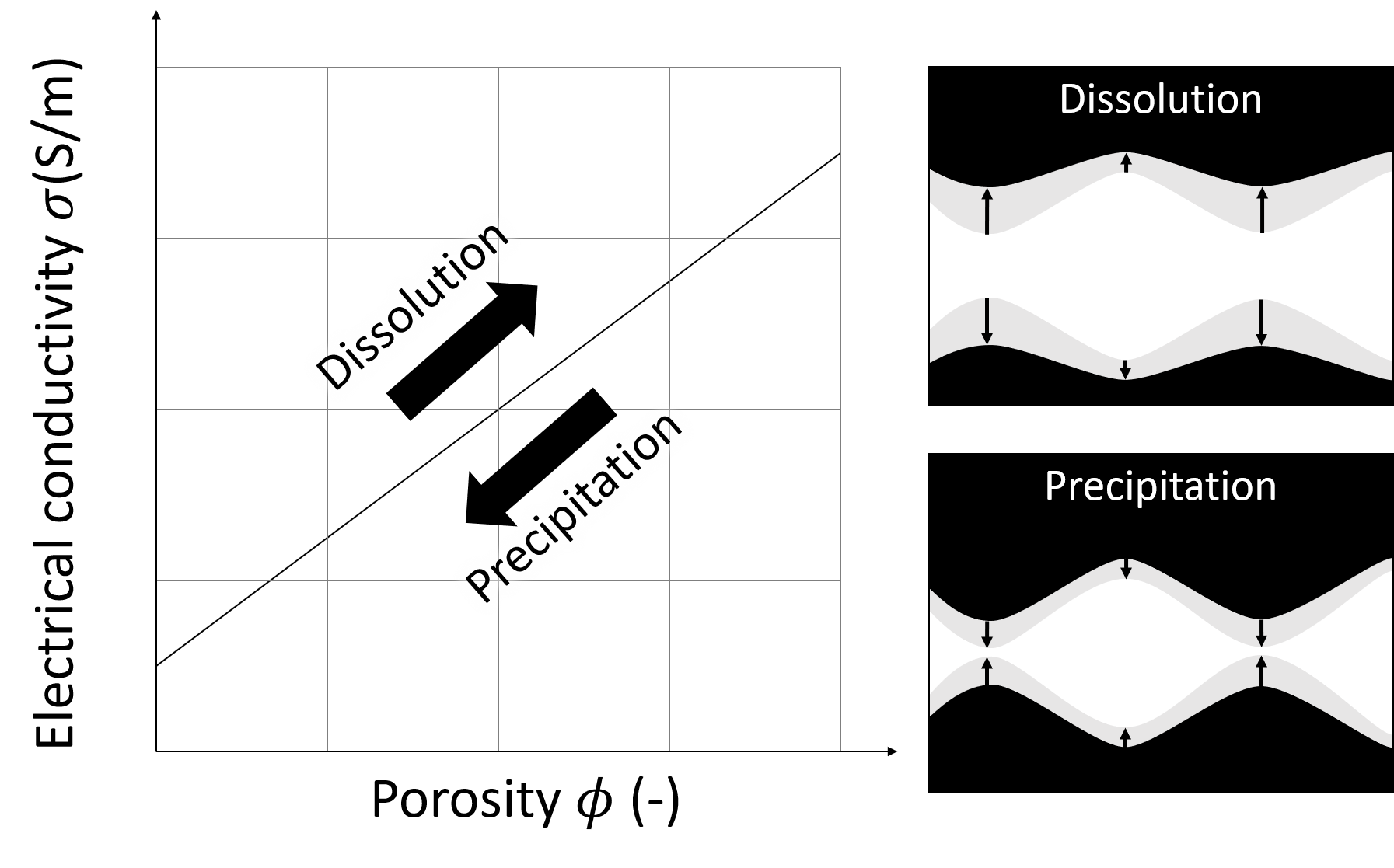}
    \caption*{Graphical abstract: a new electrical conductivity model taking into account the effect of dissolution and precipitation on the pore shape at the REV scale through a fractal-based upscaling procedure.}
\end{figure}

\newpage


\section{Introduction}

    \paragraph{}Carbonates represent a large part of the sedimentary rocks covering the Earth and carbonate aquifers store a large part of fresh water, which is a key resource for society needs. Karst aquifers are extremely complex systems because of the important chemical interactions between rock matrix and water, leading to strong chemical processes such as dissolution and precipitation. Studying these environments can benefit from the use of non-invasive tools such as the ones propose in hydrogeophysics to monitor flow and transport quantitatively \citep[e.g.,][]{hubbard-linde, binley2015}. 
    
    \paragraph{}Among the geophysical methods used for hydrological purposes in carbonate formations, electrical and electromagnetic methods have already shown their usefulness and are increasingly used \citep[e.g.,][]{chalikakis2011,revil2012,binley2015}. Electrical methods, such as direct current (DC) resistivity and induced polarization (IP), involve acquisitions with flexible configurations of electrodes in galvanic or capacitive contact with the subsurface \citep[][]{hubbard-linde}. These methods are increasingly used in different approaches to cover a larger field of applications: from samples measurements in the lab \citep[e.g.,][]{wu2010}, to measurements in one or between several boreholes \citep[e.g.,][]{daily1992} and 3D or 4D monitoring with time-lapse imaging or with permanent surveys \citep[e.g.,][]{watlet2018,saneiyan2019,mary2020}. Geophysical methods based on electromagnetic induction (EMI) consist in the deployment of electromagnetic coils in which an electric current of varying frequency is injected. Depending on the frequency range, the distance, and size of the coils for injection and reception, the depth of investigation can be highly variable \citep[][]{reynolds1998}. As for the electrical methods, EMI based methods can be deployed from the ground surface, in boreholes, and in an airborne manner \citep[e.g.,][]{paine2003}.
    
    \paragraph{}These methods enable to determine the spatial distribution of the electrical conductivity in the subsurface. They are, hence, very useful in karst-system to detect the emergence of a sinkhole, to identify infiltration area, or to map ghost-rock features \citep[e.g.,][]{jardani2006,chalikakis2011,Kaufmann2014,watlet2018}. The electrical conductivity can then be related to properties of interest for hydrogeological characterization through the use of accurate petrophysical relationships \citep[][]{binley-kemna}. In recent works, electrical conductivity models are used to characterize chemical processes between rock matrix and pore water such as dissolution and precipitation \citep[e.g.,][]{leroy2017,niu}. Indeed, geoelectrical measurements are an efficient proxy to describe pore space geometry \citep[e.g.,][]{garing,jougnot2018} and transport properties \citep[e.g.,][]{jougnot2009,jougnot2010,hamamoto2010,maineult2018}.
    
    \paragraph{}The electrical conductivity $\sigma$ (S/m) of a water saturated porous medium (e.g., carbonate rocks) is a petrophysical property related to electrical conduction in the electrolyte through the transport of charges by ions. Then, $\sigma$ is linked to pore fluid electrical conductivity $\sigma_w$ (S/m) and to porous medium microstructural properties such as porosity $\phi$ (-), pore geometry, and surface roughness. \cite{archie} proposed a widely used empirical relationship for clean (clay-free) porous media that links $\sigma$ and $\sigma_w$ to $\phi$ as follows
    \begin{equation}
        \sigma~=~\sigma_w~\phi^m,
    \end{equation}
    where $m$ (-) is the cementation exponent, defined between 1.3 and 4.4 for unconsolidated samples and for most of well-connected sedimentary rocks \citep[e.g.,][]{friedman2005}. For low pore water conductivity, porous medium electrical conductivity can also depend on a second mechanism, which can be described by the surface conductivity term $\sigma_s$ (S/m). This contribution to the overall rock electrical conductivity is caused by the presence of charged surface sites on the minerals. This causes the development of the so-called electrical double layer (EDL) with counterions (i.e., ions of the opposite charges) distributed in the Stern layer and the diffuse layer \citep[][]{hunter1981,chelidze1999,leroy2004}. Groundwater in carbonate reservoirs typically presents a conductivity comprised between 3.0$\times$10$^{-2}$~S/m and 8.0$\times$10$^{-2}$~S/m \citep[e.g.,][]{linan2009,meyerhoff2014,jeannin2016}, while carbonate rich rocks surface conductivity can range from 2.9$\times$10$^{-4}$~S/m to 1.7$\times$10$^{-2}$~S/m depending on the amount of clay \citep[][]{guichet2006,li2016,ahmed2020}. Thus, for the study of dissolution and precipitation of water saturated carbonate rocks at standard values of $\sigma_w$, the surface conductivity is generally low and can be neglected \citep[e.g.,][]{cherubini}. The small surface conductivity can nevertheless be considered as a parallel conductivity with an adjustable value \citep[e.g.,][]{waxman1968,weller2013,revil}:
    \begin{equation}
        \sigma~=~\frac{1}{F}~\sigma_w~+~\sigma_s.
        \label{sig_s}
    \end{equation}
    The formation factor $F$ (-) is thus assessed using a petrophysical law. Besides, since the late 1950's many models linking $\sigma$ to $\sigma_w$ were developed. Most of these relationships have been obtained from the effective medium theory \citep[e.g.,][]{pride1994,bussian,revil1998,ellis}, volume averaging \citep[e.g.,][]{linde2006,revil-linde}, the percolation theory \citep[e.g.,][]{broadbent,hunt}, or the cylindrical tube model \citep[e.g.,][]{pfannkuch,kennedy&herrick}. More recently, the use of fractal theory \citep[e.g.,][]{yu2001,mandelbrot} of pore size has shown good results to describe petrophysical properties among which the electrical conductivity \citep[e.g.,][]{guar-jougnot,thanh}. Meanwhile, several models have been developed to study macroscopic transport properties and chemical reactions by describing the porous matrix microscale geometry \citep[e.g.,][]{reis,guarracino,niu} and theoretical petrophysical models of electrical conductivity have been derived to relate the pore structure to transport parameters \citep[e.g.,][]{johnson,revil&cathles,glover}.
    
    \paragraph{}Permeability prediction from electrical measurements is the subject of various research studies and these models often rely on petrophysical parameters such as the tortuosity \citep[e.g.,][]{revil1998,niu}. Moreover, the use of models such as \cite{archie} and \cite{carman1939} to relate the formation factor, the porosity, and the permeability is reasonable for simple porous media such as unconsolidated packs with spherical grains, but it is less reliable for real rock samples or to study the effect of dissolution and precipitation processes. The aim of the present study is to develop a petrophysical model based on micro structural parameters, such as the tortuosity, the constrictivity \citep[i.e., parameter which is related to bottleneck effect in pores, described by][]{holzer2013}, and the Johnson length \citep[e.g.,][]{johnson,bernabe-maineult}, to express the electrical conductivity and to evaluate the role of pore structure.
    
    \paragraph{}The present manuscript is divided into three parts. We first develop equations to describe the electrical conductivity of a porous medium  with pores defined as tortuous capillaries that follow a fractal size distribution and presenting sinusoidal variations of their aperture. Then, the model is linked to other transport parameters such as permeability and ionic diffusion coefficient. In the second part, we test the model sensitivity and we compare its performance with \cite{thanh} fractal model. In the third part, we confront the model to datasets presenting an increasing complexity: first data come from synthetic unconsolidated samples, then they are taken from natural rock samples with a growing pore space intricacy. Finally, we analyze the model response to numerical simulations of dissolution and precipitation, highlighting its interest as a monitoring tool for such critical processes.


\section{Theoretical developments}
    
\begin{figure}[ht]
        \centering
        \includegraphics[width=450pt]{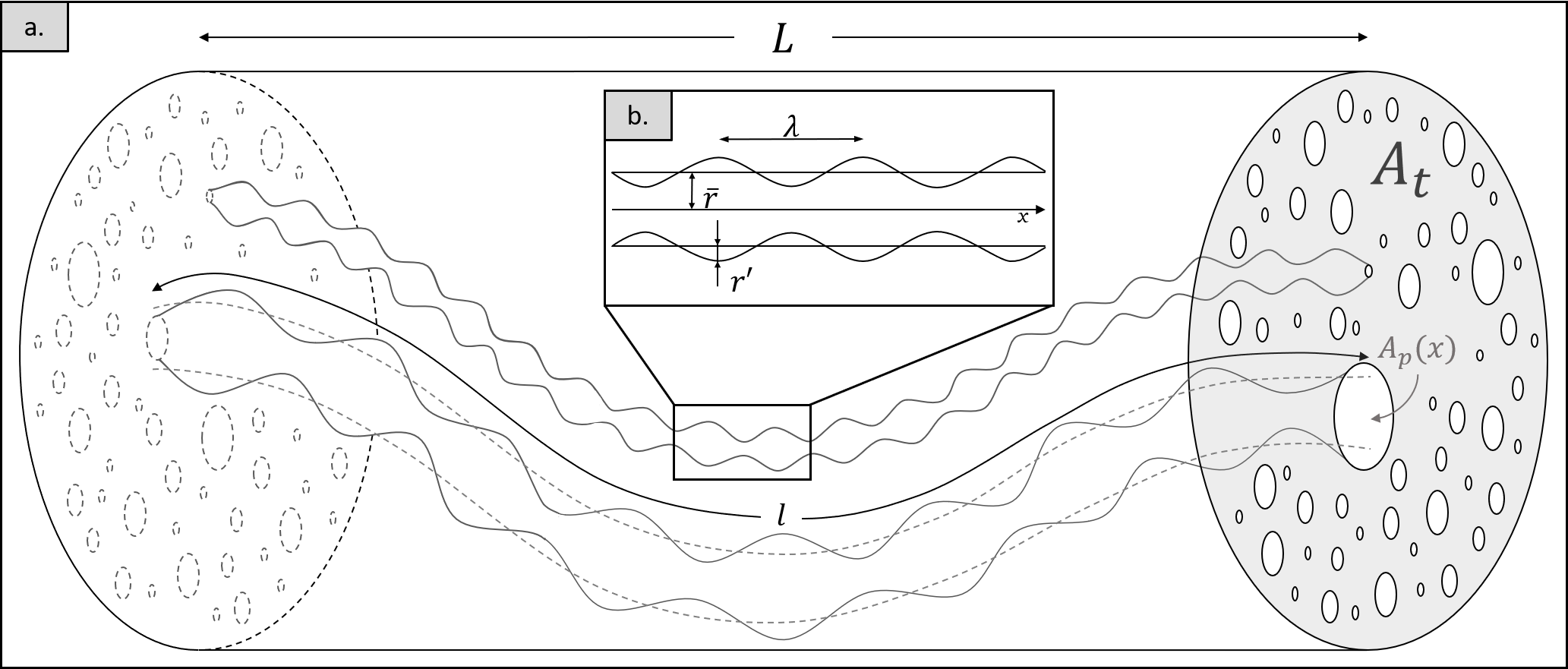}
        \caption{(a) The porous rock model is composed of a large number of sinusoidal and tortuous capillaries in the cylindrical representative elementary volume (REV). All the capillaries have the same tortuous length $l$ (m) and their radii follow a fractal distribution. (b) The considered pore geometry corresponds to the one from \cite{guarracino}: $\bar{r}$ is the average pore radius (m) while $r'$ is the amplitude of the sinusoidal fluctuation (m), and $\lambda$ is the wavelength (m).}
        \label{fig:scheme}
    \end{figure}

    \paragraph{}Based on the approach of \cite{guarracino}, we propose a model assuming a porous medium represented as a fractal distribution of equivalent tortuous capillaries in a cylindrical representative elementary volume (REV) with a radius $R$ (m) and a length $L$ (m) (Fig.~\ref{fig:scheme}a). In this model, the surface conductivity $\sigma_s$ is neglected ($\sigma_s \xrightarrow{} 0$).
        
    \subsection{Pore scale}
    
        \subsubsection{Pore geometry}
        \label{pore_geometry}
        
            \paragraph{}The porous medium is conceptualized as an equivalent bundle of capillaries. As presented in Fig.~\ref{fig:scheme}b, each tortuous pore present a varying radius $r(x)$ (m) defined with the following sinusoidal expression,
            \begin{equation}
                r(x)~=~\bar{r} + r'sin\left(\frac{2\pi}{\lambda}x\right)~=~\bar{r}\,\left(1+2a~sin\left(\frac{2\pi}{\lambda}x\right)\right),
                \label{rx}
            \end{equation}
            where $\bar{r}$ is the average pore radius (m), $r'$ the amplitude of the radius size fluctuation (m), and $\lambda$ is the wavelength (m). The parameter $a$ is the pore radius fluctuation ratio (-) defined by $a=r'/2\bar{r}$, which values range from 0 to 0.5. Note that $a~=~0$ corresponds to cylindrical pores ($r(x)~=~\bar{r}$), while $a~=~0.5$ corresponds to periodically closed pores. For each pore we define the section area $A_p(x)~=~\pi r(x)^2$ (m$^2$).
            
            \paragraph{}Most of the models found in the literature, and describing the porous medium with a fractal distribution, define a pore length scaling with pore radius \citep[e.g.,][]{yu,yu2003,guarracino,thanh}. However, in this study we consider a constant tortuous length $l$ (m) for all the pores because it reduces the number of adjustable parameters while maintaining the model accuracy. This constant tortuosity value should be interpreted as an effective macroscopic value for all tube lengths. $l$ is the length taken at the center of the capillary. Thus, the tortuosity $\tau$ (-) is also a constant for all pores and is defined as
            \begin{equation}
                \tau~=~\frac{l}{L}.
                \label{tau}
            \end{equation}
            In this case, the volume of a single pore $V_p(\bar{r})$ (m$^3$) can be computed by integrating its section area $A_p(x)$ over the tortuous length $l$:
            \begin{equation}
                V_p(\bar{r})~=~\int_{0}^{l} {\pi r(x)^2 dx}.
                \label{vp1}
            \end{equation}
            According to Eqs.~(\ref{rx}) and (\ref{tau}), and assuming that $\lambda \ll l$, volume $V_p$ defined in Eq.~(\ref{vp1}) becomes
            \begin{equation}
                V_p(\bar{r})~=~\pi \bar{r}^2 (1+2a^2) \tau L.
                \label{vp2}
            \end{equation}


        \subsubsection{Pore electrical conductivity}
            
            \paragraph{}We express electrical properties at pore scale before obtaining them for the porous medium by upscaling, because the REV can be considered as an equivalent circuit of parallel conductances, when $\sigma_s$ is neglected. The electrical conductance $\Sigma_{pore}(\bar{r})$ ($S$) of a single sinusoidal pore is defined by
            \begin{equation}
                \Sigma_{pore}(\bar{r})~=~\left(\int_{0}^{l} {\frac{1}{\sigma_w \pi r(x)^2} dx}\right)^{-1},
                \label{gpore1}
            \end{equation}
            where $\sigma_w$ (S/m) is the pore-water conductivity. Replacing Eq.~(\ref{rx}) in Eq.~(\ref{gpore1}) and assuming $\lambda\ll l$, the electrical conductance of a single pore can be expressed as
            \begin{equation}
                \Sigma_{pore}(\bar{r})~=~\frac{\sigma_w \pi \bar{r}^2 (1-4a^2)^{3/2}}{\tau L}.
                \label{gpore2}
            \end{equation}
            
            \paragraph{}Following Ohm's law, the electric voltage $\Delta$V (V) between the edges of the capillary (0 and $l$) is defined as
            \begin{equation}
                \Delta V~=~-\frac{i(\bar{r})}{\Sigma_{pore}i(\bar{r})},
            \end{equation}
            where $i(\bar{r})$ (A) is the electric current flowing through the pore that can be expressed as follows
            \begin{equation}
                i(\bar{r})~=~\frac{-\pi~\sigma_w~\bar{r}^2~(1-4a^2)^{3/2}}{\tau~L}~\Delta V.
                \label{j}
            \end{equation}
            We, thus, define the contribution to the porous medium conductivity from a single pore $\sigma_p(\bar{r}$) (S/m) by multiplying the pore conductance with a geometric factor $f_g~=~\pi~R^2/L$~(m)
            \begin{equation}
                \sigma_p(\bar{r})~=~\frac{\Sigma_{pore}(\bar{r})}{f_g}~=~\frac{\bar{r}^2 (1-4a^2)^{3/2} \sigma_w}{\tau R^2}.
            \end{equation}
            When $a~=~0$, the expression of $\sigma_p(\bar{r})$ simplifies itself as in the case of cylindrical tortuous pores developed by \cite{pfannkuch}.

    \subsection{Upscaling procedure using a fractal distribution}
    
        \paragraph{}To obtain the electrical conductivity of the porous medium at the REV scale, we need a pore size distribution. We conceptualize the porous medium by a fractal distribution of capillaries according to the notations of \cite{guarracino} and \cite{thanh}, based on the fractal theory for porous media \citep[][]{tyler,yu}
        \begin{equation}
            N(\bar{r})~=~\left(\frac{\bar{r}_{max}}{\bar{r}}\right)^{D_p},
            \label{N}
        \end{equation}
        where $N$ (-) is the number of capillaries whose average radius are equal or larger than $\bar{r}$, $D_{p}$ (-) is the fractal dimension of pore size and $\bar{r}_{max}$ (m) is the maximum average radius of pores in the REV. Fractal distributions can be used to describe objects of different Euclidean dimensions (e.g., 1 dimension for a line, 2 dimensions for a surface, and 3 dimensions for a volume). In this study, the pore size distribution is considered as a fractal distribution of capillary sections on a plane (i.e., in 2 dimensions). Therefore, the fractal dimension $D_p$ is defined from 1 to 2 \citep[among many other papers, see][]{yu,yu2003}. Nevertheless, $D_p$ is a unique parameter for each porous medium as it strongly depends on the pore size distribution. Its impact has been quantified by \cite{tyler} with a porous medium defined as a Sierpinski carpet. From the pore size distribution defined in Equation~(\ref{N}), the total number of capillaries equals to
        \begin{equation}
            N_{tot}~=~\left(\frac{\bar{r}_{max}}{\bar{r}_{min}}\right)^{D_p},
        \end{equation}
        with $\bar{r}_{min}$ (m) the minimum average radius. From Eq.~(\ref{N}), the number of radii lying between $\bar{r}$ and $\bar{r}+d\bar{r}$ is 
        \begin{equation}
            -dN~=~D_p~\bar{r}_{max}^{~D_p}~\bar{r}^{~-D_p-1}~d\bar{r},
            \label{dN}
        \end{equation}
        where $-dN$ (-) is the number of pores with an average radius comprised in the infinitesimal range between $\bar{r}$ and $\bar{r}+d\bar{r}$. The minus sign implies that the number of pores decreases when the average radius increases \citep[][]{yu2003,soldi2017,thanh}.

    
    \subsection{REV scale}
    
    \paragraph{}In the present section, we present the macroscopic properties at the REV scale obtained from the upscaling procedure.
    
        \subsubsection{Porosity}
        
            We can express the porosity $\phi$ (-) of the REV by integrating the pore volume over the fractal distribution as follows
            \begin{equation}
                \phi~=~\frac{\int_{\bar{r}_{min}}^{\bar{r}_{max}} {V_p(\bar{r})} (-dN)}{\pi R^2 L}.
                \label{phi1}
            \end{equation}
            Then, by replacing Eqs.~(\ref{dN}) and (\ref{vp2}) into Eq.~(\ref{phi1}), it yields to
            \begin{equation}
                \phi~=~\frac{(1+2a^2) \tau D_p \bar{r}_{max}^{D_p}}{R^2 (2-D_p)} (\bar{r}_{max}^{2-D_p}-\bar{r}_{min}^{2-D_p}).
                \label{phi2}
            \end{equation}
            This expression requires $2-D_p>0$, which is always true \citep[see][]{yu2001}. Note that this expression corresponds to the model of \cite{guarracino} when the tortuosity is the same for all the capillary sizes.


        \subsubsection{Electrical conductivity}
        
        \paragraph{}As defined in the Kirchhoff's current law, the electric current of the REV, $I$ (A), is the sum of the electric currents of all the capillaries when the surface conductivity is neglected. It can be obtained by integrating the electric current of each pore:
        \begin{equation}
            I~=~\int_{\bar{r}_{min}}^{\bar{r}_{max}} i(\bar{r}) (-dN).
        \end{equation}
        According to Eqs. (\ref{tau}), (\ref{j}), and (\ref{dN}), $I$ can be expressed as follows,
         \begin{equation}
            I~=~\frac{-\sigma_w \pi (1-4a^2)^{3/2} D_p \bar{r}_{max}^{D_p}}{(2-D_p)\tau L} \Delta V (\bar{r}_{max}^{2-D_p}-\bar{r}_{min}^{2-D_p}).
            \label{I1}
        \end{equation}
        The Ohm's law at the REV scale yields to
        \begin{equation}
            I~=~-\sigma^{REV} \pi R^2 \frac{\Delta V}{L},
            \label{I2}
        \end{equation}
        where $\sigma^{REV}$ is the electrical conductivity of the REV (S/m). By combining Eqs.~(\ref{I1}) and~(\ref{I2}), $\sigma^{REV}$ is expressed as
        \begin{equation}
            \sigma^{REV}~=~\frac{\sigma_w D_p \bar{r}^{D_p}_{max} (1-4a^2)^{3/2}} {R^2 \tau (2-D_p)} (\bar{r}_{max}^{2-D_p}-\bar{r}_{min}^{2-D_p}).
            \label{sigREV1}
        \end{equation}
        Finally, substituting Eq.~(\ref{phi2}) into Eq.~(\ref{sigREV1}) yields to
        \begin{equation}
            \sigma^{REV}~=~\frac{\sigma_w \phi (1-4a^2)^{3/2}} {\tau^2 (1+2a^2)}.
            \label{sigREV2}
        \end{equation}
        Note that if $a$~=~0 and $\tau$~=~1, Eq.~(\ref{sigREV2}) becomes $\sigma^{REV}~=~\sigma_w\phi$, which is the expression of Archie's law for $m$~=~1 where the porous medium is composed of a bundle of straight capillaries with no tortuosity \citep[see][]{clennell}.
        
        \paragraph{}The electrical conductivity can be rewritten depending on the tortuosity $\tau$ and on the constrictivity $f$ (-) as \begin{equation}
            \sigma^{REV}~=~\frac{\sigma_w \phi f} {\tau^2}.
        \end{equation}
        The constrictivity $f$ is thus defined as
        \begin{equation}
            f~=~\frac{(1-4a^2)^{3/2}} {(1+2a^2)}.
        \end{equation}
        The above equation highlights that the pore fluctuation ratio $a$ plays the role of the constriction factor defined by \cite{petersen1958}. Constrictivity $f$ ranges between 0 (e.g., for trapped pores) and 1 (e.g., for cylindrical pores with constant radius). As for the tortuosity $\tau$, there is no suitable method to determine constrictivity value directly from core samples, but only some mathematical expressions for ideal simplified geometries \citep[see][for a review]{holzer2013}. Therefore, very high tortuosity values \citep[e.g.,][]{niu} must be due to that in most studies the bottleneck effect is not considered.
        

        \subsubsection{Formation factor}

        \paragraph{}The model from \cite{archie} links the rock electrical conductivity to the pore water conductivity and the porosity with the cementation exponent, which is an empirical parameter. \cite{kennedy&herrick} propose to analyze electrical conductivity data using a physics-based model, which conceptualizes the porous medium with pore throats and pore bodies as in this study and defines the electrical conductivity as follows,
        \begin{equation}
            \sigma^{REV}~=~G \sigma_w \phi,
        \end{equation}
        where $G$ (-) is an explicit geometrical factor defined between 0 and 1. According to our models G can be expressed by
        \begin{equation}
            G~=~\frac{(1-4a^2)^{3/2}} {\tau^2 (1+2a^2)}~=~\frac{f} {\tau^2}.
        \end{equation}
        This geometrical factor can be called the connectedness \citep[][]{glover2015}, while the formation factor $F$ (-) is defined by
        \begin{equation}
            F~=~\frac{\sigma_w}{\sigma^{REV}}.
            \label{F1}
        \end{equation}
        Substituting Eq.~(\ref{sigREV2}) into Eq.~(\ref{F1}) yields to
        \begin{equation}
            F~=~\frac{\tau^2 (1+2a^2)}{\phi (1-4a^2)^{3/2}}~=~\frac{\tau^2}{\phi f}.
            \label{F2}
        \end{equation}
        The formation factor $F$ can also be related to the connectedness $G$ as $F~=~1/\phi G$.

       
    \subsection{Evolution of the petrophysical parameters}
    \label{evolution_of_a&tau}
    
    The formation factor defined by Eq.~(\ref{F2}) depends linearly with the inverse of porosity (1/$\phi$). However, the petrophysical parameters $a$ and $\tau$ may be dependent on porosity for certain types of rocks or during dissolution or precipitation processes. In these cases, the formation factor will show a non-linear dependence with 1/$\phi$ and can be expressed in general as
    \begin{equation}
        F(\phi) = \frac{\tau(\phi)^2 (1+2a(\phi)^2)}{\phi (1-4a(\phi)^2)^{3/2}}.
        \label{F3}
    \end{equation}

    \paragraph{}In section~\ref{conso_samples}, we test our model against different datasets from literature using logarithmic laws for the dependence of petrophysical parameters $a(\phi)$ and $\tau(\phi)$ with porosity following existing models from the literature \citep[see][for a review about the tortuosity]{ghanbarian2013}. Thus, we define $a(\phi)$ and $\tau(\phi)$ as
        \begin{equation}
            a(\phi)~=~- P_{a}\log(\phi)
            \label{a_phi}
        \end{equation}
        and
        \begin{equation}
            \tau(\phi)~=~1 - P_{\tau}\log(\phi),
            \label{tau_phi}
        \end{equation}
       where $P_a$ and $P_{\tau}$ are empirical parameters. Note that 0 and 1 (i.e., first terms in Eqs.~(\ref{a_phi}) and~(\ref{tau_phi}), respectively) correspond to the minimum values reached by $a(\phi)$ and $\tau(\phi)$ when $\phi$~=~1. Expressing tortuosity as a logarithmic function of porosity has already proven its effectiveness in the literature \citep[][]{comiti1989,ghanbarian2013,zhang2020}. However, this is, to the best of our knowledge, the first attempt to propose a constrictivity model as a function of porosity. Then, by replacing Eqs.~(\ref{a_phi}) and~(\ref{tau_phi}) in Eq.~(\ref{F3}), the expression of the proposed model for the formation factor $F$ becomes
        \begin{equation}
            F(\phi)~=~\frac{\left[1 - P_{\tau}\log(\phi)\right]^2 \left(1+2\left[ P_{a}\log(\phi)\right]^2\right)}{\phi \left(1-4\left[ P_{a}\log(\phi)\right]^2\right)^{3/2}}.
            \label{F_phi}
        \end{equation}
        Note that the model parameters from \cite{thanh}, another porous medium description following a fractal distribution of pores, also present logarithmic dependencies with the porosity $\phi$.
        
       
    \subsection{Electrical conductivity and transport parameters}
    
        \subsubsection{From electrical conductivity to permeability}
        
            \paragraph{}The electrical conductivity is a useful geophysical property to describe the pore space geometry. Here we propose to express the permeability as a function of the electrical conductivity using our model. 
            
            \paragraph{}At pore scale, \cite{sisavath2001} propose the following expression for the flow rate $Q_p(\bar{r})$ (m$^3$/s) in a single capillary:
            \begin{equation}
                Q_p(\bar{r})~=~\frac{\pi}{8}~\frac{\rho g}{\mu}~\frac{\Delta h}{l}~ 
                \left[ \int_{0}^{l}\frac{1}{r^4(x)}dx \right]^{-1}.
                \label{qp1}
            \end{equation}
            where $\rho$ is the water density (kg/m$^3$), $g$ is the standard gravity acceleration (m/s$^2$), $\mu$ is the water viscosity (Pa.s) and $\Delta h$ is the hydraulic head across the REV (m). Substituting Eq.~(\ref{rx}) in Eq.~(\ref{qp1}) and assuming $\lambda \ll l$, it yields:
            \begin{equation}
                Q_p(\bar{r})~=~\frac{\pi}{8}~\frac{\rho g}{\mu}~\frac{\Delta h}{\tau L}~\bar{r}^4 ~(1-4a^2)^{3/2}.
                \label{qp2}
            \end{equation}
            Then, the total volumetric flow rate $Q^{REV}$ (m$^3$/s) is obtained by integrating Eq.~(\ref{qp2}) over all capillaries (i.e., at the REV scale)
            \begin{equation}
						\begin{array}{r c l}
                Q^{REV} & = & \int_{\bar{r}_{min}}^{\bar{r}_{max}} Q_p(\bar{r}) (-dN) \\
                \; & = & \frac{\rho g (1-4a^2)^{3/2} D_p \bar{r}_{max}^{D_p} \pi \Delta h}{8 \mu (4-D_p) \tau L} (\bar{r}_{max}^{~4-D_p} - \bar{r}_{min}^{~4-D_p}).
            \end{array}
                \label{Q1}
						\end{equation}
            Based on Darcy's law for saturated porous media, the total volumetric flow rate can be expressed as
            \begin{equation}
                Q^{REV}~=~\pi R^2 \frac{\rho g}{\mu} k^{REV} \frac{\Delta h}{L},
                \label{Q2}
            \end{equation}
            where $k^{REV}$ is the REV permeability (m$^2$). Then, combining Eqs.~(\ref{Q1}) and~(\ref{Q2}) it yields to
            \begin{equation}
                k^{REV}~=~\frac{(1-4a^2)^{3/2} D_p \bar{r}_{max}^{D_p}}{8 R^2 (4-D_p) \tau} (\bar{r}_{max}^{~4-D_p} - \bar{r}_{min}^{~4-D_p}).
                \label{k1}
            \end{equation}
        
            \paragraph{}Considering $\bar{r}_{min} \ll \bar{r}_{max}$, Eq.~(\ref{k1}) can be simplified as
            \begin{equation}
                k^{REV}~=~\frac{(1-4a^2)^{3/2} D_p \bar{r}_{max}^4}{8 R^2 (4-D_p) \tau}.
                \label{k2}
            \end{equation}
            Using the same simplification on Eq.~(\ref{phi2}), the expression of porosity becomes
            \begin{equation}
                \phi~=~\frac{(1+2a^2) \tau D_p \bar{r}_{max}^2}{R^2 (2-D_p)}.
                \label{phi3}
            \end{equation}
            Then, combining Eqs.~(\ref{k2}) and~(\ref{phi3}) yields to
            \begin{equation}
                k^{REV}~=~\frac{2-D_p}{4-D_p}~\frac{(1-4a^2)^{3/2}}{1+2a^2}~\frac{\bar{r}_{max}^2}{8\tau^2} \phi.
                \label{k3}
            \end{equation}
            Finally, the combination of Eqs.~(\ref{F2}) and~(\ref{k3}) leads to
            \begin{equation}
                k^{REV}~=~\frac{2-D_p}{4-D_p}~\frac{\bar{r}_{max}^2}{8F}.
                \label{k4}
            \end{equation}
            Note that Eq.~(\ref{k4}) relates permeability to electrical conductivity through the formation factor (see Eq.~(\ref{F1})).
            
            \paragraph{}This expression can be linked to the model of \cite{johnson}
            \begin{equation}
                k^{REV}~=~\frac{\Lambda^2}{8F},
                \label{k5}
            \end{equation}
            where $\Lambda$ (also known as the Johnson length) is a characteristic pore size (m) of dynamically connected pores \citep[][]{banavar&schwartz,ghanbarian}. Some authors proposed theoretical relationships to determine this characteristic length $\Lambda$. While \cite{revil&cathles} or \cite{glover} link it to the average grain diameter, some other publications work on the determination of $\Lambda$ assuming a porous medium composed of cylindrical pores \citep[e.g.,][]{banavar&johnson,niu}. Considering the proposed model, $\Lambda$ can therefore be written as follows
            \begin{equation}
                \Lambda~=~\sqrt{\frac{2-D_p}{4-D_p}}~\bar{r}_{max}.
                \label{lambda1}
            \end{equation}
            

        \subsubsection{From electrical conductivity to ionic diffusion coefficient}
            
        Ionic diffusion can be described at REV scale by the Fick's law \citep[][]{fick}
        \begin{equation}
            J_t~=~D_{eff}\,\pi R^2 \frac{\Delta c}{L},
            \label{fick}
        \end{equation}
        where $J_t$ (mol/s) is the diffusive mass flow rate, $D_{eff}$ (m$^2$/s) is the effective diffusion coefficient, and $\Delta c$ (mol/m$^3$) is the solute concentration difference between the REV edges. \cite{guarracino} propose to express $D_{eff}$ as a function of the tortuosity, which, in their model, depends on the capillary size. In our model, we consider that the tortuosity is constant, thus by reproducing the same development proposed by \cite{guarracino} we obtain 
        \begin{equation}
            D_{eff}~=~D_{w} \frac{(1-4a^2)^{3/2}\phi}{(1-2a^2)\tau^2},
        \end{equation}
        which can be simplified as
        \begin{equation}
            D_{eff}~=~D_w \frac{f \phi}{\tau^2}.
            \label{Deff2}
        \end{equation}
        This last expression of $D_{eff}$ as a function of the tortuosity $\tau$ and the constrictivity $f$, allows to retrieve the same equation as \cite{vanbrakel} with both the effect of the tortuosity and the constrictivity. Replacing Eq.~(\ref{F2}) in Eq.~(\ref{Deff2}) yields to
        \begin{equation}
            D_{eff}~=~\frac{D_w}{F},
        \end{equation}
        which implies
        \begin{equation}
            F~=~\frac{\sigma_{w}}{\sigma^{REV}}~=~\frac{D_{w}}{D_{eff}}.
        \end{equation}
        This result has already been demonstrated by \cite{kyi&batchelor} and \cite{jougnot2009}, among others. It means that the formation factor can be used for both electrical conductivity or diffuse properties. This point is consistent with the fact that Ohm and Fick laws are diffusion equations, where the transport of ions take place in the same pore space. The difference lies in the fact that ionic conduction and ionic diffusion consider the electric potential gradient and the ionic concentration gradient, respectively. 


\section{Model analysis and evaluation}
        
\begin{figure}[ht]
        \centering
        \includegraphics[width=370pt]{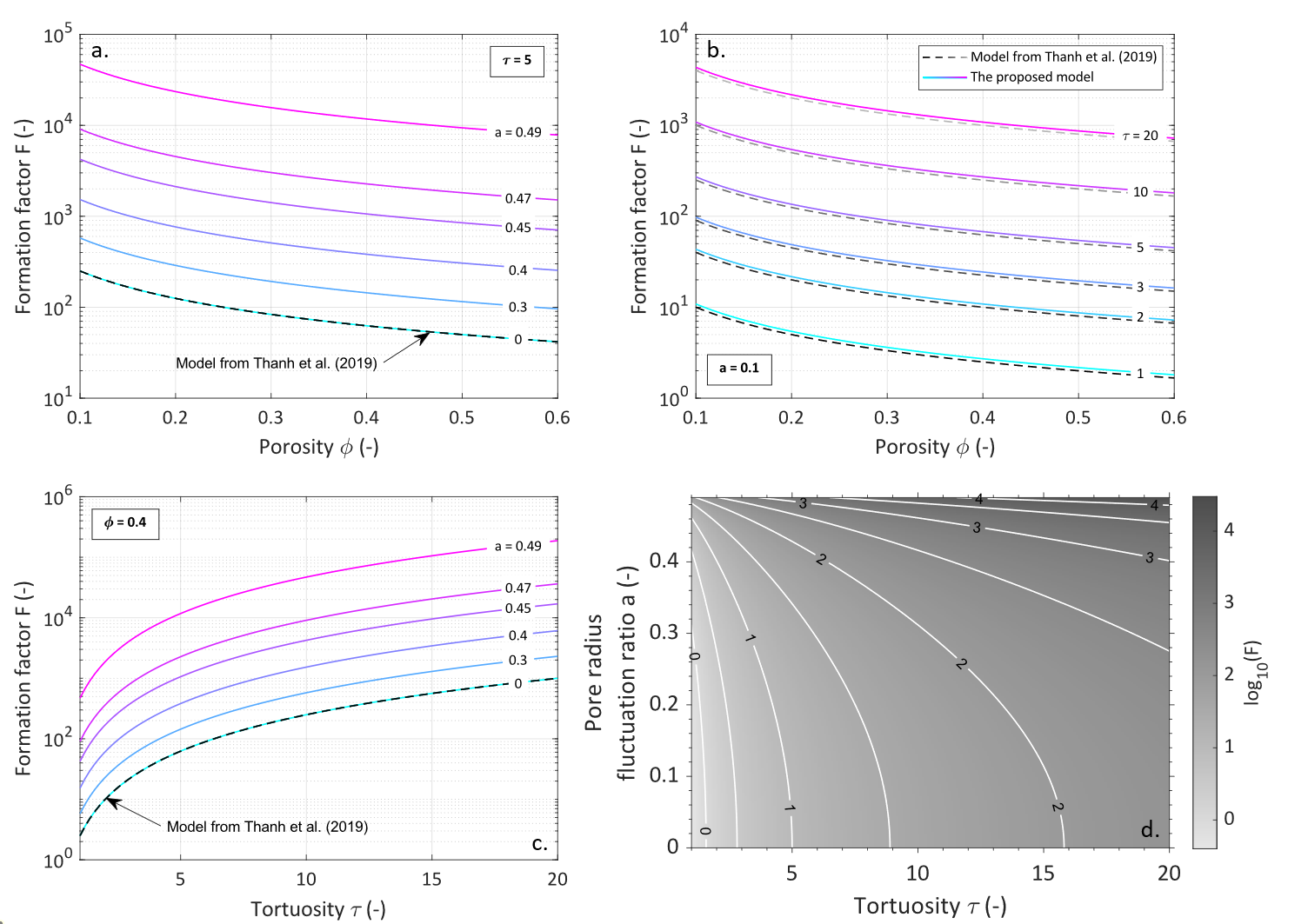}
        \caption{(a) Effect of the pore radius fluctuation ratio $a$ on the formation factor $F$, represented as a function of the porosity $\phi$. $a$ varies from 0 to 0.49, while the tortuosity $\tau$~=~5. (b) Effect of the tortuosity $\tau$ on the formation factor $F$, represented as a function of the porosity $\phi$. $\tau$ varies from 1 to 20, while the pore radius fluctuation ratio $a$~=~0.1. (c) Effect of the pore radius fluctuation ratio $a$ on the formation factor $F$, represented as a function of the tortuosity $\tau$. $a$ varies from 0 to 0.49, while the porosity $\phi$~=~0.4. (d) Comparison of the effect of parameters $a$ and $\tau$ on the formation factor $F$ for a constant porosity $\phi$~=~0.4.}
        \label{fig:sensi_FvsPhi}
    \end{figure}

    \paragraph{}Our model expresses the evolution of the formation factor $F$ as a function of the porosity $\phi$, the tortuosity $\tau$ and the constrictivity through the pore radius fluctuation ratio $a$ (Eq.~(\ref{F2})). Here we explore wide ranges of values for $a$ and $\tau$ to quantify their influence on the formation factor $F$ (Fig.~\ref{fig:sensi_FvsPhi}) and compare our model with the model from \cite{thanh}, for the fractal dimension of the tortuosity $D_{\tau}$~=~1, to appreciate the contribution of the constrictivity to the porous medium description. Figs.~\ref{fig:sensi_FvsPhi}a and~\ref{fig:sensi_FvsPhi}b show variations of $F$ as a function of the porosity $\phi$ when only $a$ or $\tau$ varies. On Fig.~\ref{fig:sensi_FvsPhi}a, parameter $a$ varies from 0 to 0.49 and tortuosity $\tau~=~5.0$. We test the case of a constant pore aperture when $a$~=~0, but we do not reach $a$~=~0.5 because this means that the pores are periodically closed (see the definition of $a$ in section \ref{pore_geometry}), and this corresponds to an infinitely resistive rock only made of non-connected porosity. On Fig.~\ref{fig:sensi_FvsPhi}b tortuosity $\tau$ varies from 1 to 20 and parameter $a~=~0.1$. $\tau$~=~1 implies straight pores (i.e., $l=L$). Fig.~\ref{fig:sensi_FvsPhi}c present variations of $F$ as a function of the tortuosity $\tau$ for different values of $a$ and a fixed porosity $\phi~=~0.4$. Fig.~\ref{fig:sensi_FvsPhi}d is the density plot of $\log_{10}(F)$ for a range of values of $a$ and $\tau$ and with a fixed porosity value $\phi~=~0.4$.
    
    \paragraph{}From the analysis of Figs.~\ref{fig:sensi_FvsPhi}a and~\ref{fig:sensi_FvsPhi}b, one can note that the formation factor decreases when porosity increases. This was expected because more the water saturated medium is porous, more its electrical conductivity is close to pore water electrical conductivity. Furthermore, we observe that both parameters $a$ and $\tau$ have a strong effect on formation factor variation ranges. Indeed, $F(\phi)$ curve can be shifted by more than 3 orders of magnitude with variations of $a$ or $\tau$ and, as expected, the formation factor increases when $\tau$ or $a$ increases. Indeed, when these parameters increase, the porous medium becomes more complex: the increase of parameter $a$ means that periodical aperture of capillaries decreases (i.e., more constrictivity), while the increase of $\tau$ means that pores become more tortuous (i.e., more tortuosity). Besides, for $a$ close to 0 (i.e., without aperture variation), curves from \cite{thanh} model are similar with the curves from our model. This is consistent with the fact that \cite{thanh} also conceptualize the porous media as a fractal distribution of capillaries. However, when $a$ increases, the curves explore very different ranges of $F$ values (Figs.~\ref{fig:sensi_FvsPhi}a and~\ref{fig:sensi_FvsPhi}c). The density plot presented on Fig.~\ref{fig:sensi_FvsPhi}d compares the effect of $a$ and $\tau$ variations for a constant porosity ($\phi~=~0.4$). It appears that for a fixed value of $\tau$, variations of $a$ have a low effect on log of $F$ values. On the contrary, variations of $\tau$ for one value of $a$ have stronger effect on log of $F$ ranges. However it should be noted that this representation can be biased because value ranges of $a$ and $\tau$ are very different. Thus, it is difficult to asses if the tortuosity $\tau$ has really much more effect on formation factor than parameter $a$. Nevertheless, it has the advantage to represent the combined effect of $a$ and $\tau$ on the formation factor.
        

\section{Results and discussion}

    \paragraph{}To assess the performance of the proposed model, we compare predicted values to datasets from the literature. References are listed in Table~\ref{tab:table_data} and ordered with a growing complexity. Indeed, data from \cite{friedman} and \cite{boleve} are taken from experiments made on unconsolidated medium. Then, \cite{garing}, \cite{revil}, and \cite{cherubini} studied natural consolidated samples from carbonate rocks and sandstone samples. Finally, \cite{niu} present numerical but dynamic data under dissolution and precipitation conditions.
    
    \begin{table}[ht]
        \centering
        \renewcommand{\arraystretch}{1.3}
        \caption{Parameters of the proposed model compared to several datasets from different sources. $\sigma_s$ is the surface conductivity (S/m) used for several comparisons out of our model application range, that is when the surface conductivity cannot be neglected, using Eq.~(\ref{sig_s}). The model parameters are adjusted with a Monte-Carlo approach, except for $a$ in \cite{niu} dataset, where $a$ is adjusted with the least square method. $\epsilon$ is the cumulative error computed in percentage, called the mean absolute percentage error (MAPE).}
        \begin{tabular}{ccccccc}
          \hline
          \multirow{2}*{Sample} & \multirow{2}*{$a$} & \multirow{2}*{$\tau$} & $\sigma_s$ & $\epsilon$ & Studied & \multirow{2}*{Source} \\
                 &     &        & (S/m) & (\%) & function & \\
          \hline
          S1a            & 0.004         & 1.035         & 2.25 $\times10^{-4}$ & 7.78  & \multirow{6}*{$\sigma(\sigma_w)$} & \multirow{6}*{\cite{boleve}} \\
          S2             & 0.008         & 1.062         & 1.45 $\times10^{-4}$ & 7.47  \\
          S3             & 0.006         & 1.050         & 0.80 $\times10^{-4}$ & 7.09  \\
          S4             & 0.006         & 1.051         & 0.50 $\times10^{-4}$ & 9.21  \\
          S5             & 0.012         & 1.093         & 0.25 $\times10^{-4}$ & 10.22 \\
          S6             & 0.008         & 1.062         & 0.60 $\times10^{-4}$ & 9.21  \\
          \hline
          Glass  & 0.022         & 1.174         & -    & 0.32  & \multirow{3}*{$F(\phi)$} & \multirow{3}*{\cite{friedman}} \\
          Sand           & 0.026         & 1.212         & -    & 0.60  \\
          Tuff           & 0.020         & 1.159         & -    & 1.63  \\
          \hline
          FS$\,^{a}$     & 0.146 - 0.309 & 1.365 - 1.773 & -     & 22.62 & $F(\phi)$ & \cite{revil} \\
          \hline
          L1, L2         & 0.113         & 1.901         & 7.24 $\times10^{-4}$ & 9.24  & $\sigma(\sigma_w)$ & \cite{cherubini} \\
          \hline
          Inter          & 0.077 - 0.217 & 1.846 - 3.399 & -    & 26.67 & \multirow{3}*{$F(\phi)$} & \multirow{3}*{\cite{garing}} \\
          Multi          & 0.172 - 0.345 & 1.915 - 2.839 & -    & 9.64  \\
          Vuggy          & 0.068 - 0.109 & 5.632 - 8.411 & -    & 19.68 \\
          \hline
          D.T.lim$\,^{b}$      & 0.078 - 0.393 & 1.786         & -    & 0.05 & \multirow{4}*{$F(\phi)$} & \multirow{4}*{\cite{niu}} \\
          D.R.lim$\,^{c}$      & 0.315 - 0.393 & 1.786         & -    & 0.08 \\
          P.T.lim$\,^{d}$    & 0.167 - 0.466 & 1.335         & -    & 0.12 \\
          P.R.lim$\,^{e}$    & 0.160 - 0.309 & 1.320         & -    & 0.04 \\
          \hline
          \multicolumn{6}{l}{$^{a}$~FS: Fontainebleau sandstones} \\
          \multicolumn{6}{l}{$^{b}$~D.T.lim: Dissolution transport-limited} \\
          \multicolumn{6}{l}{$^{c}$~D.R.lim: Dissolution reaction-limited} \\
          \multicolumn{6}{l}{$^{d}$~P.T.lim: Precipitation transport-limited} \\
          \multicolumn{6}{l}{$^{e}$~P.R.lim: Precipitation reaction-limited}
        \end{tabular}
        \label{tab:table_data}
    \end{table}
    
    \paragraph{}For each dataset, the adjusted parameters of the proposed model are listed in Table~\ref{tab:table_data}. Values have been determined using a Monte-Carlo inversion of Eqs.~(\ref{sigREV2}) and~(\ref{F2}) which express the porous medium electrical conductivity $\sigma^{REV}$ as a function of the pore water conductivity $\sigma_w$ and the formation factor $F$ as a function of the porosity $\phi$, respectively. An additional term $\sigma_s$ for the surface conductivity (S/m) is used to fit the data out of the application range of the proposed model. That is for low values of pore-water conductivity, when the surface conductivity cannot be neglected. As the proposed model is intended to be mostly used on carbonate rocks, which are known to have low surface conductivity, this physical parameter has not been taken into account in the theoretical development of the expression of the electrical conductivity of the porous medium. However, it can be added considering a parallel model \citep[e.g.,][]{waxman1968,borner1991}:
    \begin{equation}
        \sigma^{REV}~=~\frac{1}{F}\sigma_w + \sigma_s.
    \end{equation}
    A more advanced approach of parallel model is proposed by \cite{thanh} including the contribution of the surface conductance in the overall capillary bundle electrical conductivity.
    
    \paragraph{}Table~\ref{tab:table_data} lists also the computed error $\epsilon$ of the adjusted model. In statistics $\epsilon$ is called the mean absolute percentage error (MAPE). It is expressed in percent and defined as follow:
     \begin{equation}
         \epsilon~=~\frac{1}{N^d}\left( \sum^{N^d}_{i=1}\left|\frac{P^m_i - P^d_i}{P^d_i}\right|\right)\times100,
         \label{mape}
     \end{equation}
    where $N^d$, $P^d$, and $P^m$ refer to the number of data, the electrical property from data, and the electrical property from the model, respectively. This type of error has been chosen to compare the ability of the model to reproduce the experimental values for all the datasets even if they are expressed as $\sigma^{REV}(\sigma_w)$ or as $F(\phi)$.
    

    \subsection{Testing the model on unconsolidated media}
        
        \paragraph{}The proposed model is first confronted to datasets from \cite{friedman} and \cite{boleve} obtained for unconsolidated samples. In these tests, only one set of parameters $a$, $\tau$, and $\sigma_s$ (when needed) is adjusted to fit with each dataset.
        
\begin{figure}[ht]
            \centering
            \includegraphics[width=370pt]{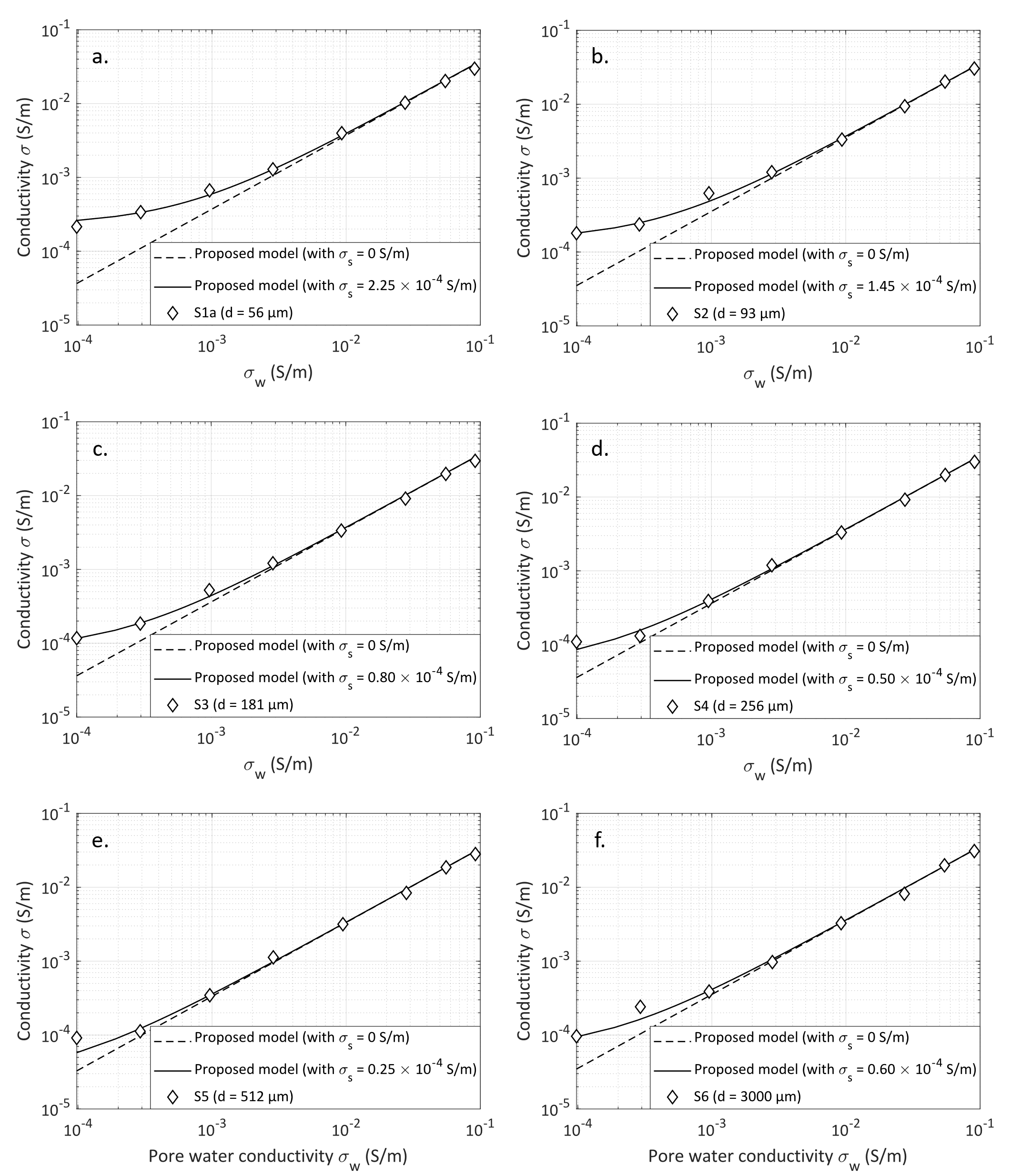}
            \caption{Electrical conductivity of different samples of glass beads (grains sizes are 56, 93, 181, 256, 512, and \unit{3000}{\micro\meter} for samples S1a, S2, S3, S4, S5, and S6, respectively) versus the fluid electrical conductivity for a constant porosity $\phi~=~40~\%$. The datasets are from \cite{boleve} and best fit parameters are given in Table~\ref{tab:table_data}.}
            \label{fig:boleve}
        \end{figure}

        \paragraph{}\cite{boleve} measured the electrical conductivity of glass beads samples for different values of the pore water conductivity $\sigma_w$ from $10^{-4}$~S/m to $10^{-1}$~S/m on S1a, S2, S3, S4, S5, and S6 (see Fig.~\ref{fig:boleve}). For all samples, \cite{boleve} reported a constant porosity of 40~\% while grain diameters are comprised between \unit{56}{\micro\meter} for S1a and \unit{3000}{\micro\meter} for S6 (see Fig.~\ref{fig:boleve} for more details). 
        
        \paragraph{}Table~\ref{tab:table_data} shows that the adjusted model parameters $a$ and $\tau$ have rather similar values for all the samples from \cite{boleve}. This can be explained by the fact that all samples have the same pore geometry but scaled at different size. Indeed, for homogeneous samples of glass beads, the beads space arrangement is quite independent of the spheres size. Therefore the pore network of all samples presents similar properties such as tortuosity \citep[see also the discussion in][]{guar-jougnot} and constrictivity, which directly depends on parameters $a$ and $\tau$. Moreover, $a$ and $\tau$ are close to their minimum limits (i.e., $a$~=~0 and $\tau$~=~1). This is due to the simple pore space geometry created by samples made of homogeneous glass spheres. This explains the model good fit for straight capillaries \citep[i.e.,][]{thanh}. However, it can be noticed that surface conductivity decreases while grain diameter increases. This is not a surprise considering that for the same volume, samples of smaller beads have a larger specific surface than samples of bigger beads \citep[see, for example, the discussion in][]{glov-dery}.
        
\begin{figure}[ht]
            \centering
            \includegraphics[width=200pt]{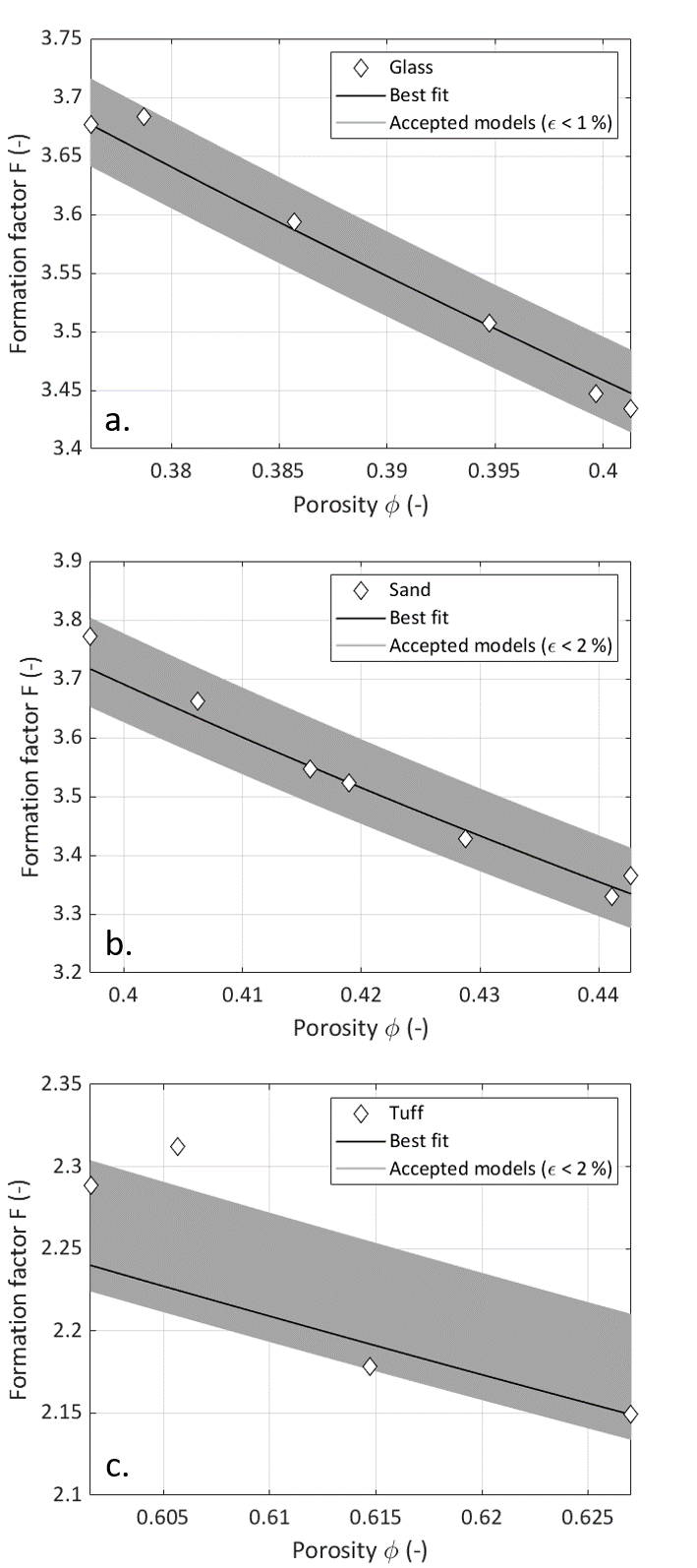}
            \caption{Formation factor of samples with different porosity values: (a) glass beads, (b) sand, and (c) tuff grains \citep[][]{friedman}. The sets of adjusted model parameters $a$ and $\tau$ are given in Table~\ref{tab:table_data}.}
            \label{fig:friedman}
        \end{figure}

        \paragraph{}\cite{friedman} determined the formation factor $F$ for samples of glass beads, sand, and tuff grains with different values of porosity $\phi$ (see Fig.~\ref{fig:friedman}). As the model best fit is determined with a Monte Carlo approach, accepted models are also plotted on Fig.~\ref{fig:friedman}. The acceptance criterion is defined individually for each dataset and corresponds to a certain value of the MAPE $\epsilon$. For samples from \cite{friedman}, this criterion is fixed at $\epsilon<$~1 or~2~\%. These are low values, meaning a really good fit from the model.
        
        \paragraph{}We observe from Table~\ref{tab:table_data} that $a$ and $\tau$ values are close to each other for all samples from \cite{friedman}. However, these parameters have higher values than for \cite{boleve} dataset. This comes from the fact that some complexity is added in the dataset from \cite{friedman}. Indeed, the samples of \cite{friedman} combine particles of different sizes. In this case, smaller grains can fill the voids left by bigger grains. This grains arrangement decreases the medium porosity but increases its tortuosity and constrictivity. Furthermore, sand and tuff grains have rougher surface and are less spherical than glass beads. This explains the misfit increase between data and model for glass, sand, and tuff samples \citep[][]{friedman}. Nevertheless, even for tuff grains, the misfit between data and model is still very low compared to the computed MAPE from \cite{boleve} samples. This is due to that for \cite{boleve}, a wide range of pore water conductivity values is explored and thus electrical properties vary much more (over 3 orders of magnitude) than in the case of \cite{friedman}.
        
        
    \subsection{Testing the model on consolidated rock samples}
    \label{conso_samples}
        
        \paragraph{}In this section, we test our model against datasets of \cite{garing}, \cite{revil}, and \cite{cherubini}. They study carbonate samples from the reef unit of Ses Sitjoles site (from Mallorca), Fontainebleau sandstones, and two Estaillades limestones (rodolith packstones), respectively.
        
        \begin{figure}[ht]
            \centering
            \includegraphics[width=200pt]{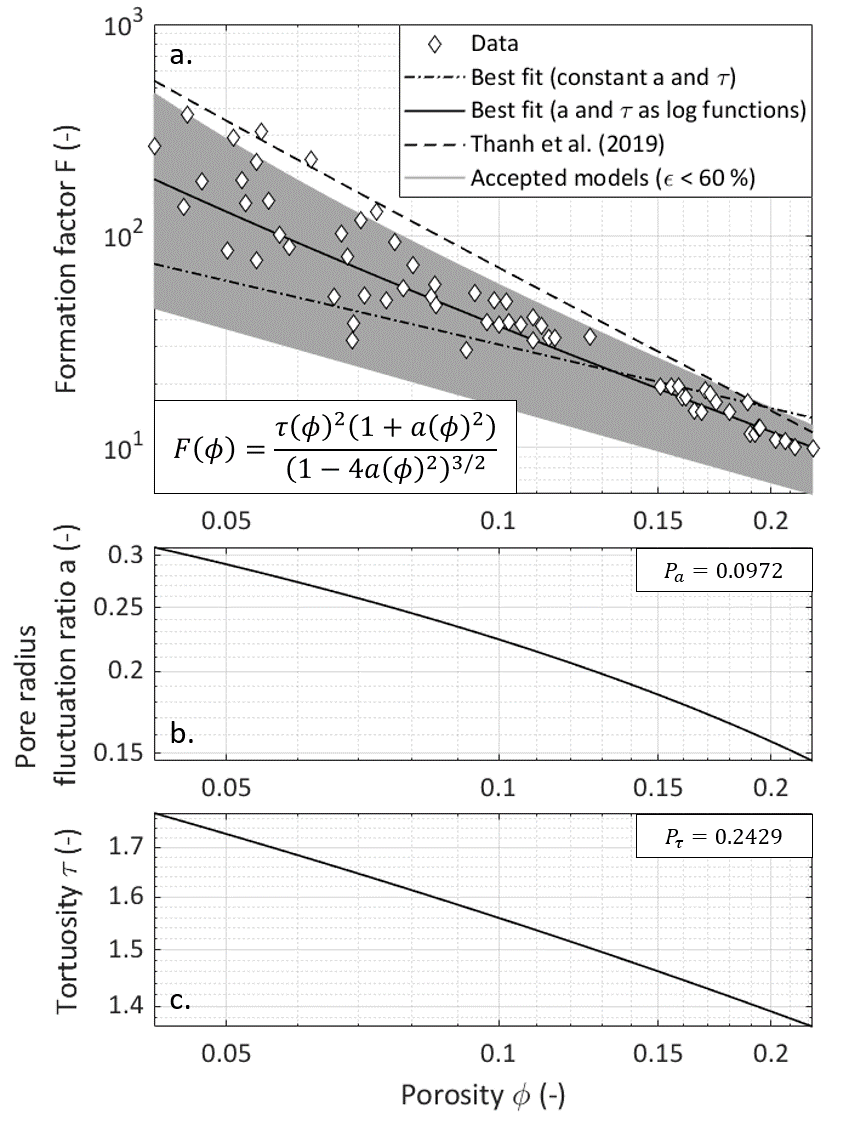}
            \caption{(a) Formation factors of a set of Fontainebleau sandstones versus porosity \citep[the dataset is from][]{revil}. Model parameters are given in Table~\ref{tab:table_data}. 
            (b) and (c) $a$ and $\tau$ are defined as logarithmic functions of the porosity $\phi$.}
            \label{fig:revil}
        \end{figure}
        
        \paragraph{}\cite{revil} obtained a wide range of formation factor values for a large set of Fontainebleau sandstone core samples over a large range of porosity. We first test our model with constant values of $a$ and $\tau$, but we observed that the model could not fit data (see Fig.~\ref{fig:revil}a). This can be explained by the fact that this dataset is composed of numerous sandstone samples presenting a wide range of porosity values (from 0.045 to 0.22). Therefore, we consider that the samples have distinct pore geometry which is describable by  $a(\phi)$ and $\tau(\phi)$ distributions, presented in section \ref{evolution_of_a&tau} and plotted on Figs.~\ref{fig:revil}b and \ref{fig:revil}c. We observe that parameters $a$ and $\tau$ logarithmic evolution with the porosity $\phi$ are physically plausible as lower porosity can reflect more complex medium geometries (i.e., more constrictive and more tortuous), described with higher values of $a$ and $\tau$. On Fig.~\ref{fig:revil}a, we also plot the model from \cite{thanh}. Even if the curve presents a slope similar to dataset, it overestimates the formation factor.
        
        \paragraph{}Despite a quite wide dispersion of the formation factor data for the lowest porosities, it appears that the proposed model is well adjusted to the dataset. Indeed, the proposed model MAPE $\epsilon~=~22.62~$\% (see Table~\ref{tab:table_data}), while $\epsilon~=~89.63~$\% for the model from \cite{thanh}. Note that the relatively high MAPE value comes from the large spread of the formation factor values. Thus, it seems that taking into account the constrictivity of the porous medium in addition to the tortuosity highly improves modeling.
        
        \begin{figure}[ht]
            \centering
            \includegraphics[width=250pt]{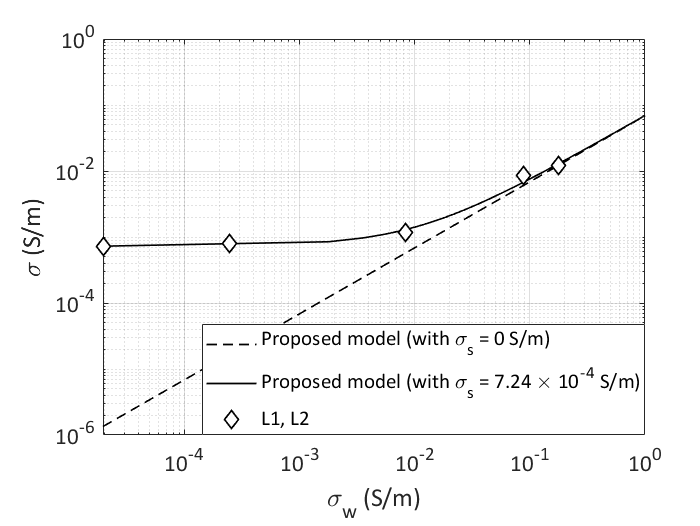}
            \caption{Electrical conductivity of two limestones (L$_1$ and L$_2$) versus water electrical conductivity. The dataset is from \cite{cherubini} and model parameters are given in Table~\ref{tab:table_data}.}
            \label{fig:cherubini}
        \end{figure}
        
        \paragraph{}Fig.~\ref{fig:cherubini} shows the dependence of electrical conductivity with pore-water electrical conductivity for two limestone samples (named L1 and L2) from \cite{cherubini}. Table~\ref{tab:table_data} shows that model parameters and surface conductivity values are larger than for the unconsolidated samples from \cite{boleve} and \cite{friedman}. This is explained by the fact that natural rock samples can present a more complex geometry and larger specific surface area than glass beads samples. \cite{cherubini} predict the surface electrical conductivity with the model from \cite{revil}. They obtain $\sigma_s~=~7.0\times10^{-4}$~S/m, which is very close to the value obtained in this study. Furthermore, the computed errors for the dataset of \cite{boleve} and for these limestones are close to each other. This test illustrates that even for more complex porous media, the proposed model has still a good data resolution.
        
        \begin{figure}[ht]
            \centering
            \includegraphics[width=200pt]{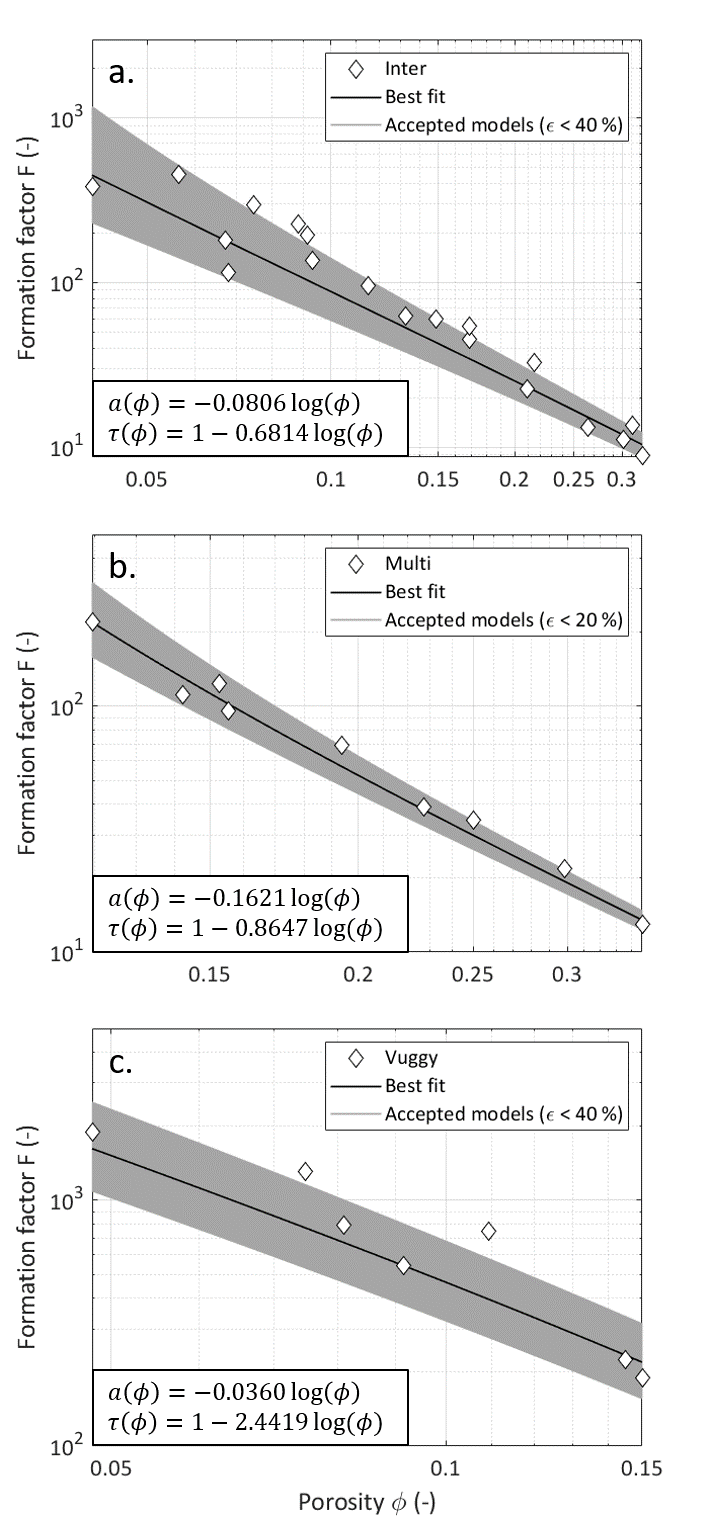}
            \caption{Formation factors of a set of carbonate rocks classified by pore types versus porosity. The dataset is from \cite{garing}. The model parameters are given in Table~\ref{tab:table_data}. The proposed model parameters are considered to be logarithmic functions of $\phi$. (a) ``inter'' stands for samples with intergranular pores (b) ``multi'' gathers samples with multiple porosity types: intergranular, moldic, and vuggy. (c) ``vuggy'' represents samples with vugular porosity.}
            \label{fig:garing}
        \end{figure}
        
        \paragraph{}\cite{garing} conducted X-ray microtomography measurements on carbonate samples to classify them by pore types and thus they highlight three groups:
        \begin{enumerate}
            \item ``Inter'' samples present intergranular pores. This pore shape is quite comparable with sandstones porosity type.
            \item ``Multi'' samples hold multiple porosity types: intergranular, moldic, and vugular. Microtomograms of ``multi'' samples show small but well connected pores. The analyze conducted by \cite{garing} revealed that for samples with smaller porosity, pores are smaller on average, but still numerous and well connected, even for a reduced microporosity.
            \item ``Vuggy'' samples possess vugular porosity. Microtomography highlights the presence of few vugs badly connected, which are less numerous for samples of lower porosity.
        \end{enumerate}
        Fig.~\ref{fig:garing} presents the results of formation factor computation versus porosity. As for the dataset from \cite{revil}, the proposed model is adjusted using Eq.~(\ref{F_phi}), which considers that model parameters are logarithmic functions of porosity. Despite some dispersion for ``inter'' and ``vuggy'' samples (i.e., the acceptance criterion $\epsilon<40~$\%), the model explains well the data for all porosity types and present low MAPE values (Table~\ref{tab:table_data}).
        
        \paragraph{}The analysis of parameters $P_a$ and $P_{\tau}$ reveals that they present consistent values for the different porosity types. Indeed, for ``vuggy'' samples, $P_a$ is small while $P_{\tau}$ is high (Fig.~\ref{fig:garing}c). According to Eqs.~(\ref{a_phi}) and~(\ref{tau_phi}), these values lead to low and high values for $a$ and $\tau$, respectively (Table~\ref{tab:table_data}), and this is consistent with the microtomography analysis from \cite{garing}. Indeed, since these samples present large vugs badly connected (i.e., few microporosity), the microstructure is very tortuous but pores are not constricted. Furthermore, for samples with lower porosity, vugs are still present in ``vuggy'' samples, but they are less numerous, which leads to a microstructure even more tortuous, but nearly as constrictive as for the more porous samples. Moreover, for ``inter'' and ``multi'' samples (Figs.~\ref{fig:garing}a and~\ref{fig:garing}b), $P_a$ and $P_{\tau}$ are closer in value to the parameters of the sandstone samples from \cite{revil} than to the parameters of ``vuggy'' samples because they have, among other types for the ``multi'' samples, intergranular porosity. Note that higher $P_{\tau}$ value can be attributed to the more complex structure of carbonate minerals compared to sandstone samples. Furthermore, the high value of $P_a$ for the ``multi'' samples can be explained with the microtomography observations from \cite{garing}. Indeed, constrictivity increases a lot for samples with lower porosity because microporosity is reduced while there are less molds and vugs. Consequently, we conclude that this detailed analysis of model parameters help us to retrieve some characteristic features of the pore space from electrical conductivity measurement.
        

    \subsection{Electrical conductivity monitoring of precipitation and dissolution processes}
    \label{diss_precip}
        
        \paragraph{}In this section, we consider the numerical datasets from \cite{niu}. These authors conduct numerical simulations of dissolution and precipitation reactions on digital representations of microstructural images. They simulate the dissolution of a carbonate mudstone sample and the precipitation of a sample of loosely packed ooids. For the carbonate mudstone sample, the pore space image is obtained from a microtomography scan while the ooids sample is a synthetic compression of sparsely distributed spherical particles \citep[][]{niu2018}. The carbonate mudstone sample has an initial porosity of 13~\% and the ooids sample has an initial porosity of 30.2~\%. 
        
        \paragraph{}In numerical simulations, the main hypothesis of \cite{niu} is that fluid transport is advection dominated. Then, under this condition, they studied two limiting cases for both dissolution and precipitation: the transport-limited case and the reaction-limited case \citep[][]{nunes}. In the transport‐limited case, the reaction at the solid‐liquid interface is limited by the diffusion of reactants to and from the solid surface. In the reaction‐limited case, the reaction is limited by the reaction rate at the solid‐liquid interface. 
        
\begin{figure}[ht]
            \centering
            \includegraphics[width=400pt]{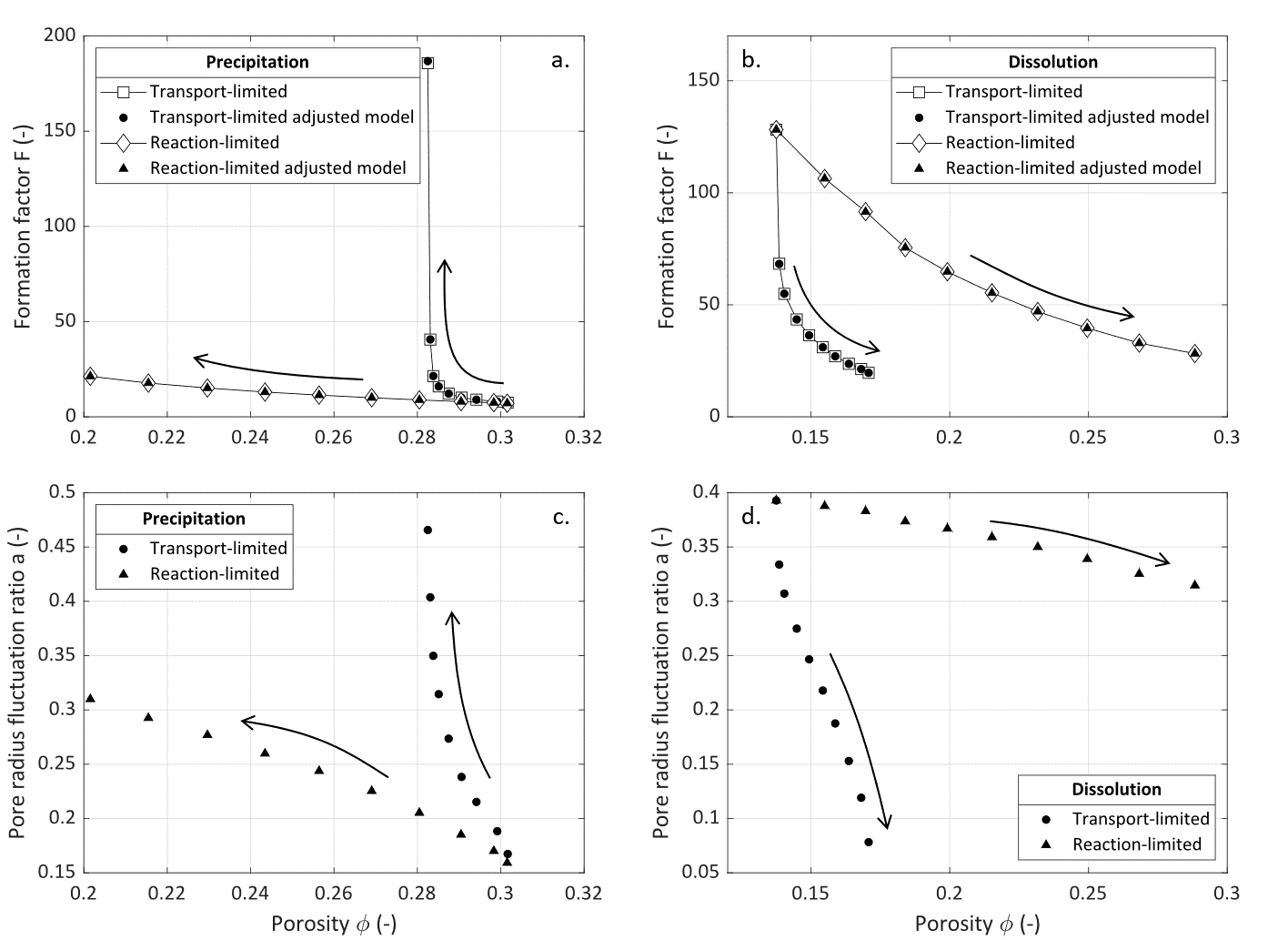}
            \caption{Electrical simulation results for two limiting cases (tansport-limited and reaction-limited) of calcite precipitation and dissolution from \cite{niu}. Arrows indicate the direction of dissolution or precipitation process evolution. (a) and (b) Formation factor versus porosity obtained by \cite{niu} simulations and compared with the adjusted model for precipitation and dissolution, respectively. Model parameters are given in Table~\ref{tab:table_data}: the tortuosity $\tau$ is considered to be constant while the pore radius fluctuation ratio $a$ is the only adjustable parameter of the model. (c) and (d) Evolution of the parameter $a$ which increases when the porosity decreases.}
            \label{fig:niu_formfactor}
        \end{figure}

\begin{figure}[ht]
            \centering
            \includegraphics[width=400pt]{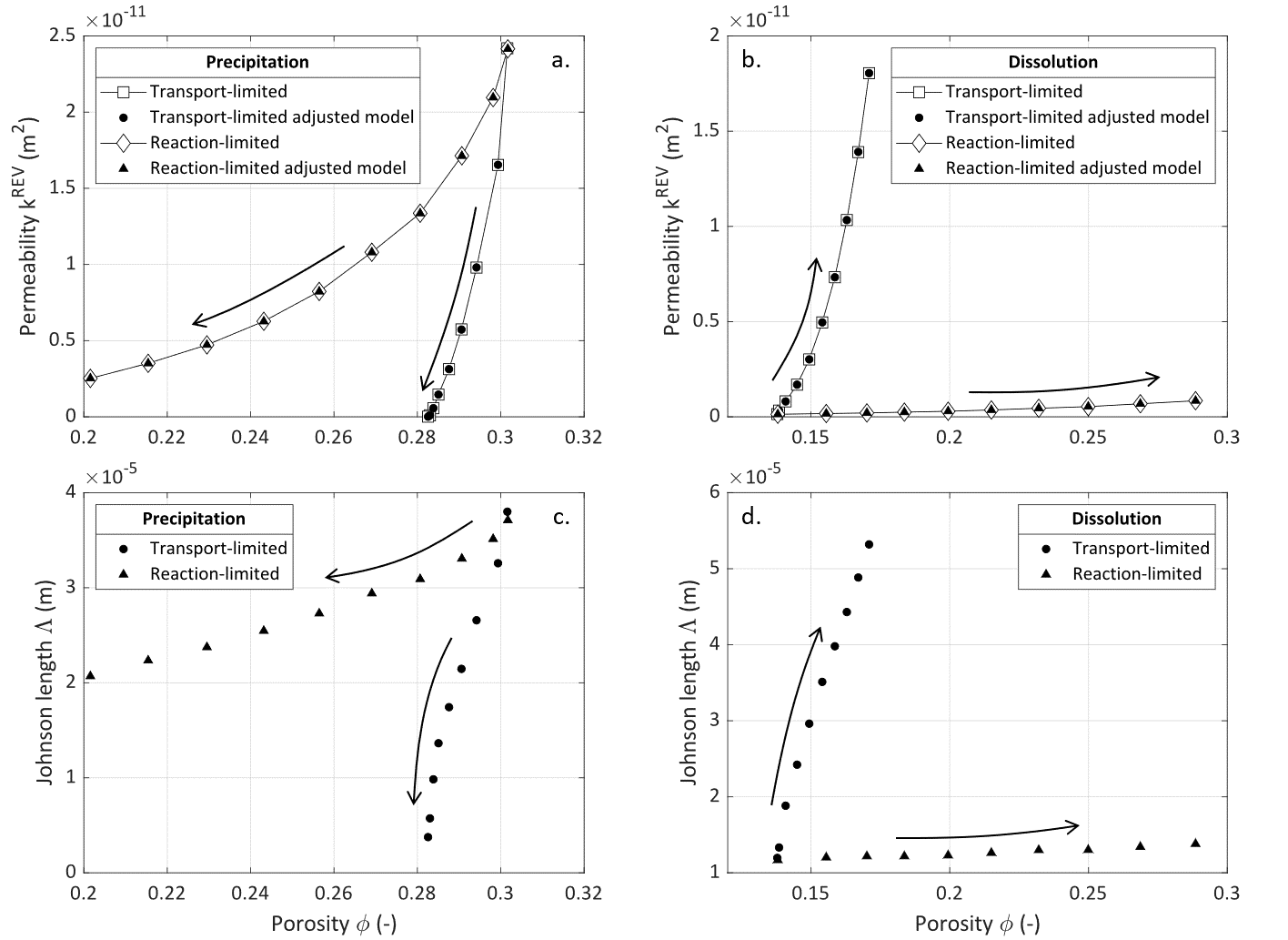}
            \caption{Fluid flow properties simulation results of two limiting cases (tansport-limited and reaction-limited) of calcite precipitation and dissolution from \cite{niu}. Arrows indicate the direction of dissolution or precipitation process evolution. (a) and (b) Permeability versus porosity obtained by \cite{niu} simulations and compared with the proposed model for precipitation and dissolution, respectively. The model parameters are given in Tables~\ref{tab:table_data} and~\ref{tab:table_perm}: as values of the tortuosity $\tau$ and the pore radius fluctuation ratio $a$ are reused from the formation factor modeling, the Johnson length $\Lambda$ is the only adjustable parameter of the model. (c) and (d) Evolution of the parameter $\Lambda$ which increases with porosity.}
            \label{fig:niu_perm}
        \end{figure}

        \paragraph{}Their results are presented in Figs.~\ref{fig:niu_formfactor}a, b, and \ref{fig:niu_perm}a, b. It appears that for both precipitation and dissolution, the transport-limited case influences the most electrical and fluid flow properties. Indeed, it can be seen in Fig.~\ref{fig:niu_formfactor}a that for reaction-limited precipitation, a 10~\% decrease in porosity leads to an increase in the formation factor from 7.5 to 20, while for transport-limited precipitation, the formation factor reaches 140 for a porosity decrease of less than 2~\%. In case of dissolution (Fig.~\ref{fig:niu_formfactor}b), for a similar decrease of the formation factor, porosity increases only by 3~\% in the transport-limited case, while it has to increase by 15~\% in the reaction-limited case. The same observations can be made on Figs.~\ref{fig:niu_perm}a and~\ref{fig:niu_perm}b. The variations of permeability are much greater in the transport-limited case than in the reaction-limited case for a lower porosity variation.
        
        \paragraph{}The authors of this study justify the shapes of $F$ and $k^{REV}$ curves with their observations on the digital representations of the microstructural evolution. In the reaction-limited case they observe that precipitated or dissolved minerals are uniformly distributed over grain surfaces. This consequently barely affects electrical and fluid flow properties. On the contrary, in the transport-limited case, dissolution and precipitation mainly occur in some specific areas where fluid velocity is high. This significantly modifies electrical and fluid flow patterns. In the case of transport-limited precipitation, new particles accumulate in pore throats while minerals are preferentially dissolved in the already well opened channels.
        
        \paragraph{}To adjust the proposed model to the data, a first set of parameters $a$ and $\tau$ has been determined at the initial state with the Monte-Carlo approach. Then, only parameter $a$ is adjusted with the model to each new data point using the least square method. We consider that only $a$ is affected by dissolution and precipitation because these processes mostly affect the pore shape. Indeed, the results of the pore network modeling developed by \cite{steinwinder2019} to simulate the impact of pore‐scale alterations by dissolution and precipitation on permeability, show that pore throats are important parameters to take into account. However, for the proposed model of this study, the assumption that only $a$ varies requires slow processes of dissolution and precipitation in order to keep the cylindrical shape of pores \citep[][]{guarracino}. Besides, we fit parameter $a$ at each time step rather than using the logarithmic law since it lacks physical meaning to explain this parameter time evolution.
        
        \paragraph{}\cite{niu} computed the hydraulic tortuosity $\tau_h$ (-) from the simulated fluid velocity field for all of their data. They found nearly constant values defined between 1 and 2. As we computed $\tau$~=~1.3 and $\tau$~=~1.8 in precipitation and dissolution, respectively, these results are within the predicted range of the simulated data from \cite{niu} and confirm our hypothesis that constrictivity is the pore feature most impacted by dissolution and precipitation processes. However, to interpret the evolution of the formation factor during dissolution ($F$ decreases) and precipitation ($F$ increases), \cite{niu} computed the electrical tortuosity ($\tau_e~=~F\phi$, no unit) of their porous medium and obtained high values (from 2 to 200), varying over 1 to 2 decades for the transport-limited cases. These overestimated values of the medium tortuosity highlight that constrictivity and the bottleneck effect should not be neglected to evaluate how dissolution and precipitation processes affect the pore structure.
        
        \paragraph{}Figs.~\ref{fig:niu_formfactor}a~and~\ref{fig:niu_formfactor}b show that in each case the proposed model accurately fits the data with computed errors lower than 1~\% (see Table~\ref{tab:table_data}). As presented on Figs.~\ref{fig:niu_formfactor}c and~\ref{fig:niu_formfactor}d, parameter $a$ follows monotonous variations: it decreases when the porosity increases. We define $a$ as the ratio of the sinusoidal pore aperture $r'$ over the mean pore radius $\bar{r}$ (see the definition of $a$ in section \ref{pore_geometry}). When $a$ increases, it can be caused by the increase of $r'$, which involves a stronger periodical constriction of the pore aperture, and/or by the decrease of the mean pore radius $\bar{r}$. On the contrary, when $a$ decreases, it implies the increase of $\bar{r}$ and/or the decrease of $r'$, which lead to thicker pores with smoother pore walls, respectively. These variations are consistent with the fact that precipitation and dissolution affect the pore geometry through the sample. In case of precipitation pore throats shrink while they are enlarged with dissolution. It can also be observed on Figs.~\ref{fig:niu_formfactor}c and~\ref{fig:niu_formfactor}d that $a$ shows stronger variations in the transport-limited case than in the reaction-limited case. This is consistent with the fact that transport-limited reactions occur in localized areas which will strongly affect the pore properties.
        
        \paragraph{}The relation between the permeability $k^{REV}$ and the Johnson length $\Lambda$ is obtained by combining Eqs.~(\ref{k3}) and~(\ref{lambda1}):
        \begin{equation}
            k^{REV}~=~\Lambda^2~\frac{(1-4a^2)^{3/2}}{1+2a^2}~\frac{\phi}{8\tau^2}
            \label{k6}
        \end{equation}
        The values of parameters $a$ and $\tau$ are taken from the adjusted models of the formation factor. Thus, the Johnson length $\Lambda$ is the only adjustable parameter to fit the data and is fitted with the least square method. Values are given in Tables~\ref{tab:table_data} and~\ref{tab:table_perm}.
        
        \begin{table}[ht]
            \centering
            \renewcommand{\arraystretch}{1.3}
            \caption{Values of the Johnson length $\Lambda$ and of the MAPE $\epsilon$ (defined in Eq.~(\ref{mape})) for the modeling of permeability versus porosity for the samples from \cite{niu}. $\Lambda$ is adjusted with the least square method. Sample names are defined in Table~\ref{tab:table_data}.}
            \begin{tabular}{ccc}
              \hline
              \multirow{2}*{Sample} & $\Lambda$ & $\epsilon$ \\
               & ($10^{-5}$ m) & (\%) \\
              \hline
              Dissolution transport-limited   & 1.196 - 5.320 & 0.10 \\
              Dissolution reaction-limited   & 1.167 - 1.380 & 0.03 \\
              Precipitation transport-limited & 0.376 - 3.800 & 0.24 \\
              Precipitation reaction-limited & 2.069 - 3.711 & 0.06 \\
              \hline
            \end{tabular}
            \label{tab:table_perm}
        \end{table}
        
\begin{figure}[ht]
            \centering
            \includegraphics[width=350pt]{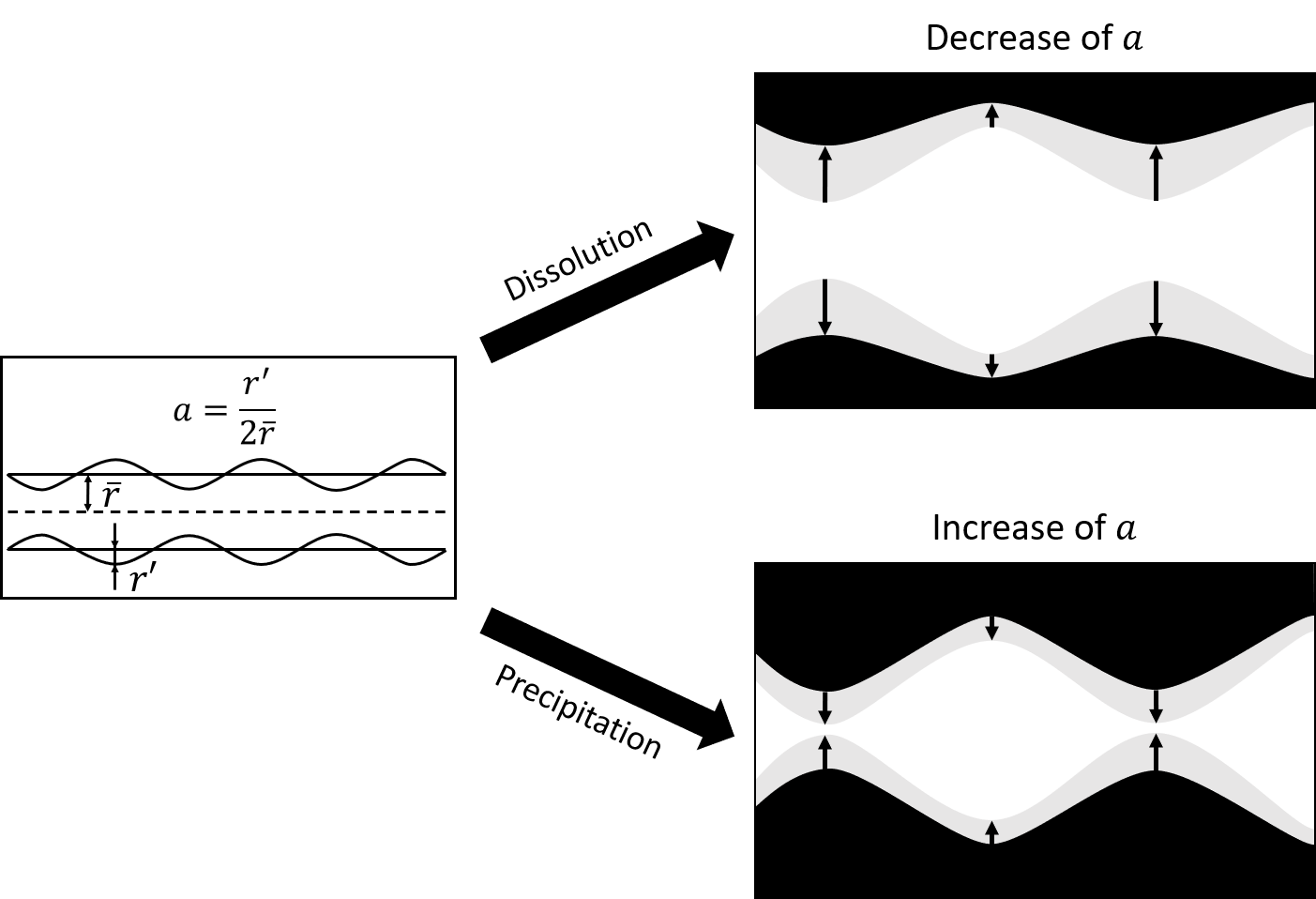}
            \caption{The pore radius fluctuation ratio $a$ is the model parameter which is updated during dissolution and precipitation reactions. Under precipitation $a$ increases, hence the pore aperture varies more. On the contrary $a$ decreases under dissolution and hence the pore becomes smoother.}
            \label{fig:evol_a}
        \end{figure}

        \paragraph{}On Figs.~\ref{fig:niu_perm}a and~\ref{fig:niu_perm}b, one can observe that for each case the model accurately fits the data with computed errors lower than 1~\% (see Table~\ref{tab:table_perm}). As presented on Figs.~\ref{fig:niu_perm}c and~\ref{fig:niu_perm}d, parameter $\Lambda$ follows monotonous variations: it increases with porosity. It can also be observed that $\Lambda$ varies more in the transport-limited case than in the reaction-limited case. For the reaction-limited dissolution it is even nearly constant. \cite{niu} found Johnson lengths with the same order of magnitude and with similar variations except in the case of transport-limited precipitation where their values do not follow a monotonous behavior for low porosity values. Either way, \cite{niu} interpret the Jonhson length as an effective radius of their porous medium which shows monotonous variations during precipitation ($\Lambda$ decreases) and dissolution ($\Lambda$ increases). In the proposed model, the parameter $a$ describes how dissolution and precipitation processes affect the shape of the pore radius (i.e., its constrictivity) while $\Lambda$ is linked to the fractal distribution of pore size $D_p$ and to the maximum average radius $\bar{r}_{max}$. As we suppose dissolution and precipitation slow enough not to interfere with the pore size distribution, $D_p$ remains constant for each sample. On the contrary, when dissolution or precipitations occurs, it is expected for the pores to grow or to shrink, respectively. Therefore, The monotonous variations of $\Lambda$ highlight the increase or decrease of $\bar{r}_{max}$ during dissolution or precipitation, respectively. This result is in accordance with the variations of $a$ which can impact $\bar{r}$. Consequently, we describe the pore space evolution during dissolution and precipitation as illustrated in Fig.~\ref{fig:evol_a}. Indeed, the decrease of $a$ is caused by dissolution, which enlarges the pore and flattens its pore walls. On the contrary, precipitation affects the pores by increasing $a$, which means that pores shrink and become more periodically constricted because of $r'$ increase. We thus believe that this interpretation of the electrical conductivity measurement is an important issue for future research on the impact of dissolution and precipitation on the pore shape.
        
\section{Conclusion}

     \paragraph{}In the present work we express the electrical conductivity and the formation factor of the porous medium in terms of effective petrophysical parameters such as the tortuosity and the constrictivity. The model is based on the conceptualization of the pore space as a fractal cumulative distribution of tortuous capillaries with a sinusoidal variation of their radius (i.e., periodical pore throats). By means of an upscaling procedure, we link the electrical conductivity to transport parameters such as permeability and ionic diffusion coefficient.
     
     \paragraph{}The proposed model successfully predicts electrical conductivity and formation factor of unconsolidated samples and natural consolidated rock samples. For datasets of sandstones or carbonates with large range of porosity values, we set that $a$ and $\tau$ follow logarithmic functions of $\phi$. These empirical relations allow the model to accurately fit the datasets. On one hand, for the sandstone samples, the prediction fits much better than previously published models, while on the other hand, the model parameter analysis shows strong agreement with the porosity types description thanks to X-ray microtomography investigations on carbonate samples.
     
     \paragraph{}Even if our model is designed for porous media in which the surface conductivity can be neglected, it is possible to take it into account at very low salinity. We do not express it as a function of the pore structure parameters, but determine its value empirically. The comparison of its value with the one found for other models on the same datasets shows that this approach is consistent and reasonable for the purpose of this model.
     
     \paragraph{}The model is finally compared to a numerical dataset from simulations of dissolution and precipitation reactions on digital representations of microstructural images. The model can reproduce structural changes linked to these processes by only adjusting a single parameter related to the medium constrictivity: the pore radius fluctuation ratio $a$. We observe that this parameter follows monotonous variations under dissolution or precipitation conditions that makes it a good witness of these chemical processes effect on the pore structure.
        
    \paragraph{}We believe that the present study contributes to a better understanding of the links between the electrical conductivity measurement, the pore space characteristics and the evolution of the microstructural properties of the porous medium subjected to dissolution and precipitation processes, therefore enhancing the possibility of using hydrogeophysical tools for the study of carbonate hydrosystems. In the future, we will extend this approach to partially saturated conditions and include these new petrophysical models in an integrated hydrogeophysical approach to better understand hydrosystems in the critical zone.
    
\section{Acknowledgments}
The authors warmly thank Qifei Niu and the other anonymous reviewer for the constructive comments that helped to greatly improve the manuscript quality. The authors wish to thank the editor for the effective editing process. The authors strongly thank the financial support of the CNRS INSU EC2CO program for funding the STARTREK (Système péTrophysique de cAractéRisation du Transport Réactif en miliEu Karstique) project. The authors are also very thankful for the wise comments of Roger Guérin.
\section{Notations}
    
    \begin{center}
    \renewcommand{\arraystretch}{1.2}
    \tablefirsthead{
        \hline
        Parameter & Definition & Unit \\
        \hline}
        
    \tablehead{
        \multicolumn{3}{l}{\small\sl continued from previous page} \\
        \hline
        Parameter & Definition & Unit \\
        \hline}
        
    \tabletail{
        \hline
        \multicolumn{3}{r}{\small\sl continued on next page}\\}
    
    \tablelasttail{\hline}
    
    \begin{supertabular}{p{2cm} p{8cm} p{5cm}}
      $L$ & REV length & m \\
      $R$ & REV radius & m \\
      $r$ & pore radius & m \\
      $\bar{r}$ & average pore radius & m \\
      $r'$ & amplitude of the radius size fluctuation & m \\
      $\lambda$ & wavelength & m \\
      $a$ & pore radius fluctuation ratio & - \\
      $A_p$ & REV section area & m$^2$ \\
      $l$ & tortuous length & m \\
      $\tau$ & tortuosity & - \\
      $V_p(\bar{r})$ & volume of a single pore & m$^3$ \\
      $\Sigma_{pore}(\bar{r})$ & electrical conductance of a single pore & $S$ \\
      $\rho_w$ & pore water electrical resistivity & $\Omega$.m \\
      $\sigma_w$ & pore water electrical conductivity & S/m \\
      $\Delta V$ & electric voltage & V \\
      $i(\bar{r})$ & electric current flowing through a single pore & A\\
      $\sigma_p(\bar{r})$ & contribution to the porous medium conductivity from a single pore & S/m \\
      $f_g$ & geometric factor & m \\
      $N(\bar{r})$ & number of pores of radius equal or larger than $\bar{r}$ & - \\
      $\bar{r}_{max}$ & maximum average pore radius & m \\
      $D_p$ & fractal dimension of pore size & - \\
      $N_{tot}$ & total number of pores & - \\
      $\bar{r}_{min}$ & minimum average pore radius & m \\
      $\phi$ & porosity & - \\
      $I$ & REV electric current & A \\
      $\sigma^{REV}$ & REV electrical conductivity & S/m \\
      $f$ & constrictivity & - \\
      $G$ & connectedness & - \\
      $F$ & formation factor & - \\
      $Q_p(\bar{r})$ & flow rate of a single pore & m$^3$/s \\
      $\rho$ & density of water & kg/m$^3$ \\
      $g$ & standard gravity & m/s$^2$ \\
      $\mu$ & water viscosity & Pa.s \\
      $\Delta h$ & hydraulic charge across the REV & m \\
      $Q^{REV}$ & total volumetric flow rate & m$^3$/s \\
      $k^{REV}$ & REV permeability & m$^2$ \\
      $\Lambda$ & Johnson length & m \\
      $J_t$ & diffusive solute mass flow rate & mol/s \\
      $D_w$ & water diffusion coefficient & m$^2$/s\\
      $D_{eff}$ & effective diffusion coefficient & m$^2$/s \\
      $\Delta$c & solute concentration differences & mol/m$^3$ \\
      $m$ & cementation exponent & - \\
      $\sigma_s$ & surface conductivity & S/m \\
      $\epsilon$ & mean absolute percentage error & \% \\
      $N^d$ & number of data & - \\
      $P^m_i$ & electrical property from the model & - \\
      $P^d_i$ & electrical property from the data & - \\
      $P_a$ & coefficient to define $a(\phi)$ & - \\
      $P_{\tau}$ & coefficient to define $\tau(\phi)$ & - \\
      $\tau_h$ & hydraulic tortuosity & - \\
      $\tau_e$ & electrical tortuosity & - \\
      \hline
    \end{supertabular}
    \end{center}
    
\bibliography{article_tortuous_v9}
\bibliographystyle{plainnat}
\end{onehalfspace}
\end{document}